 \documentclass[nohyper,12pt,letterpaper]{JHEP}
 \usepackage{epsfig}
 \newenvironment{Footnote}{\footnote\bgroup}{\egroup}
 \title{Regulating Eternal Inflation}
 \author{T.\,Banks\\
 Department of Physics and SCIPP\\
 University of California, Santa Cruz, CA 95064\\
 E-mail: \email{banks@scipp.ucsc.edu}\\
 {\it and}\\
 Department of Physics and NHETC, Rutgers University\\
 Piscataway, NJ 08540}

 \author{M.\,Johnson\\
 Department of Physics and SCIPP\\
 University of California, Santa Cruz, CA 95064\\
 E-mail: \email{mjohnson@physics.ucsc.edu}}

 \abstract{We present an interpretation of the physics of space-times
 undergoing {\it eternal inflation} by repeated nucleation of bubbles.
 In many cases the physics can be interpreted in terms of the quantum
 mechanics of a system with a finite number of states.
 If this interpretation is correct, the conventional picture of these
 space-times is misleading.}
 \received{} \accepted{}
 \preprint{hep-th{0512141}\\\\ \\}
 
\begin{document}

 \section{\bf Introduction}

 {\it Eternal Inflation}\cite{guthvillindepjs} or
 {\it The Self-Reproducing Universe}\cite{l} is
 one of the most confusing ideas to come out of the attempt to marry
 general relativity with quantum field theory.  There are two classes
 of eternally inflating universe.  The simplest model in the first
 class consists of a scalar field with a potential like that in Fig.
 1, coupled to gravity.  The false minimum has a positive value
 and we will consider all three possibilities for the sign of the
 energy density in the true minimum $V_T = \sigma |V_T|,\ \sigma
 = \pm 1,\ 0$.   Bubbles of true vacuum nucleate inside the
 false\cite{cdl},
 but as long as the false vacuum is dS space,
 the interior of any bubble is causally disconnected from the
 causal diamonds of a large class of observers\footnote{Observer $=$ large localized quantum
system with many semi-classical observables.  An observer is well
described by finite volume, cut-off quantum field theory, with a
large volume in cut-off units.  The semi-classical observables are
averages of local fields over large volumes.  By their nature,
observers in this sense are always massive, and follow time-like
trajectories through space-time.}. Additional bubbles
 can nucleate inside these causal diamonds.   It is generally
 claimed that any observer will eventually experience a tunneling to
 the true vacuum, but there is much confusion about how to define
 the late time asymptotics of the system.   One draws Penrose
 diagrams with a future space-like ${\cal I}_+$ and asserts that it
 has a fractal nature, divided between regions that asymptote to $T$
 and $F$.   Depending on one's choice of finite time slices, one can
 assert that at late times the overwhelming majority of space is
 still inflating, whence the name.

\EPSFIGURE{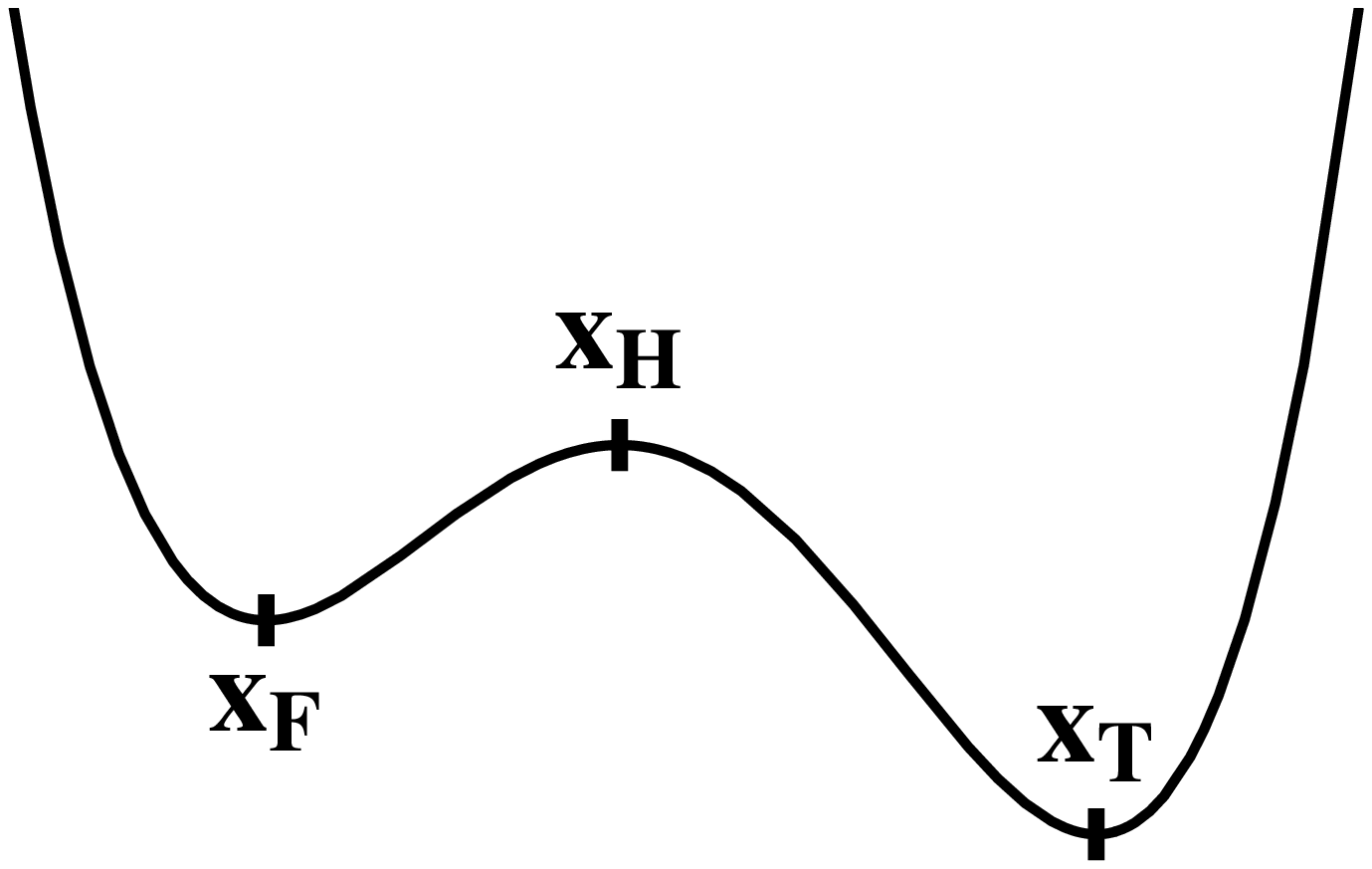,width=75mm}{The Potential for Vacuum Tunneling,
$V(x_F )$ is Positive and $V(x_T )$ Can Have Either Sign, or
Vanish.\label{hemipic}}



 The second class of eternally inflating models relies on quantum
 fluctuations in a slow-roll regime\cite{l}.  The fluctuations drive the
 field back up the potential in some regions of space.  Our
 understanding of these models in terms of holographic and thermal
 physics is much less complete, and we will not refer to them further in this
 paper.

 We will argue that if $\sigma = \pm 1$ there is another
 interpretation of the Coleman-DeLuccia\cite{cdl} tunneling process, in terms
 of a system with a finite number of states.   In this view, there
 is no fractal Penrose diagram, and no time slicing ambiguity.   The
 details of the physics depend on the sign of $\sigma$.   In the
 particularly interesting case where $\sigma$ is negative and
 $|V_T| \gg |V_F|$ (both are of course well below the Planck scale)
 we argue that it is extremely improbable for an observer to
 actually view the tunneling event
 before it is destroyed by other processes which take place in dS
 space.  The Crunch of Joy foretold by Coleman and DeLuccia is
 replaced by a more prosaic, if no less grim, demise.
 The case $\sigma = 0$ appears to describe a system with an
 infinite number of states, but the limit is delicate and will be
 discussed in more detail below.

 The next section of this paper (very briefly) reviews our
 understanding of the properties of a
 theory of stable dS space\cite{tbds}.  We emphasize the existence of
 two Hamiltonians, $H$ and $P^0$, relevant for describing the physics
 seen by a single time-like observer. $P^0$ is useful for the description
 of a small subset of low (compared to the maximal black hole mass) energy
 localized states.   We then turn to the case of tunneling between
 two dS spaces.   We recapitulate the argument\cite{heretics}
  that there is an interpretation of the
 CDL instanton transition in a Hilbert space with a large but finite
 number of states.   Most of the states resemble the dS vacuum with
 smaller c.c., but there is a small subsystem which resembles the
 other dS vacuum.   Transitions back and forth occur, and obey the
 law of detailed balance.  Recall that any given observer sees
 precisely such a sequence of transitions.  In our interpretation,
 eternal inflation is (in this case) described by a Hilbert space
 for only a single observer.  Different, causally disconnected
 observers, correspond to complementary descriptions of the same
 Hilbert space, using different time evolution operators\footnote{In the case of CDL instanton
 geometries, one of the causally disconnected observers explores only a factor space of the full
 Hilbert space, with the complementary factor in a fixed state.}.  The
 evolution operators of two causally disconnected observers do not
 commute at any time.

 Finally, we argue that this mathematical description is practically
 fictitious when one dS vacuum has very large radius.
 There is no operational way to measure the consequences of
 tunneling.   For large dS radius in the true vacuum, nucleation of macroscopic black
 holes on top of the observer occurs at a rate which is
 superexponentially\footnote{We use the term super-exponential to refer
 to two processes whose time scales behave as $e^{(RM_P)^{a_i}}$ where
 $a_1 - a_2$ is of order one.   When $RM_P > 10^2$ the longer time is
 essentially the same number measured in Planck units as it is
 in units of the shorter time scale.} faster than tunneling to the
 other vacuum.   Quantum fluctuations of the largest
 macroscopic apparatus that is allowed in dS space, also set in on
 time scales superexponentially shorter than the vacuum tunneling
 rate.

 The next section generalizes these observations to the case where
 $V_T$ is negative, while $V_F$ remains positive.  When $V_F \ll
 |V_T|$, we argue that the nomenclature $T,F$ is misleading.
 The system
 has a finite number of states, most of which resemble the dS
 vacuum.  Transitions to the CDL crunch are rare events, which
 are extremely unlikely to be observed by local observers.  In the opposite
 limit, most of the states of the system are associated with the
 Crunch, and we must find a more complete description of this
 singular space-time before we can assess whether the instanton
 corresponds to a process in a sensible quantum system.

 Finally, we investigate the case where $V_F > V_T = 0$.  This is a
 singular limit, with an infinite number of states, and we must
 carefully distinguish which subclass of states we keep in the
 limiting theory.

The title of this paper is explained by the fact that we replace the
infinite fractal web of conventional eternal inflation by a finite
system.   In the conclusions, the reader will also find a remark
that may clarify the sense in which the present work is a regulator
for the conventional picture.   This remark is based on the
observation of \cite{nightmare}, that the entropy of dS space allows
for of order $(RM_P)^{1/2}$ commuting copies of the field theoretic
degrees of freedom in a single horizon volume.   In \cite{nightmare}
it was suggested that this could be organized into a regulated
version of the global coordinate picture of dS space in quantum
field theory.   The current paper may be providing a regulated
version of conventional eternal inflation, in much the same sense.

 In addition to this remark, the concluding section discusses
  implications of our considerations
 for the String Landscape, and for the theory of stable dS space.
 Throughout the paper, we will work in four dimensions and let $M_P$
 stand for the Planck mass and $m_P$ the reduced Planck mass.

 \section{The decay of the dominant vacuum}

 \subsection{Stable dS space}

 Let us briefly review the description of stable dS space\cite{tbds}, restricting
 our attention to four space-time dimensions. The
 system has a finite number of states\footnote{This claim is based on the
 covariant entropy bound\cite{FSB}.  R. Bousso first made the hypothesis\cite{rb} that
 the CEB for a causal diamond implied a finite number of ``degrees of freedom" for that system.
 Fischler and one of the authors\cite{tbfolly}\cite{willy}, argued that this implied a finite number of
 quantum states for a stable dS space.} , close to the exponential of
 the Gibbons Hawking (GH) entropy, $S_{GH} = \pi R^2 M_P^2$.
 To describe the physics of a
 localized static observer, we need two Hamiltonians, $H$ and $P_0$.
 The first is the ``real" Hamiltonian of the system.   It has a
 highly degenerate spectrum, below $T_{dS} = {1\over 2\pi R}$, with
 level density $e^{ - S_{GH}}$.   Geometrically, these states live
 within a Planck distance of the cosmological horizon.  They
 constitute the de Sitter vacuum ensemble.

 The Poincare Hamiltonian $P_0$ has commutator
 \begin{equation}
[H, P_0] \sim {1\over R} P_0.
\end{equation}
Its eigenspaces with
 eigenvalue $<< {RM_P^2}$ are approximately conserved by the
 time evolution generated by $H$, and
 correspond to states localized near the origin of static
 coordinates. For black holes, the Poincare eigenvalue is the mass
 parameter in the Schwarzschild metric:
\begin{equation}
ds^2 = - (1 - {2 M \over {M_P^2 r}} + r^2/R^2) dt^2 + {dr^2 \over
{(1 - {2 M \over {M_P^2 r}} + r^2/R^2)}} + r^2 d\Omega^2 .
\end{equation}
All of the states with non-zero Poincare eigenvalue have smaller
entropy than the dS vacuum, and they decay back to it. In fact, in
the semiclassical approximation for the gravitational field, the
Poincare eigenvalue is just the entropy deficit of the corresponding
eigen-space, relative to the dS vacuum. This surprising connection
is necessary for the consistency of the description of dS space as a
thermal distribution of Poincare eigenstates, which is a consequence
of quantum field theory in curved space-time.   It can also be
verified by inspection of the black hole entropy formula, for those
states which are semiclassical black holes.

In the limit $R \rightarrow \infty$, the quantum theory of dS space
approaches a theory appropriate to asymptotically flat space. We get
the flat space theory by focussing on localizable states with
Poincare eigenvalues $\ll RM_P^2$, and considering {\it approximate
scattering amplitudes} of those states.  The approximate S matrix
$S_R$ is only approximately unitary, because scattering takes place
in a thermal background whose source is the dS vacuum ensemble.
However, it approaches a unitary operator as $R \rightarrow \infty$,
because the states on the horizon decouple from the localized states
and the Gibbons-Hawking temperature goes to zero. This limiting
scattering matrix is the full set of observables for the Poincare
invariant limiting theory. If the hypothesis of Cosmological
Supersymmetry Breaking (CSB) is correct, the limit is in fact
super-Poincare invariant. We will find one attractive consequence of
this hypothesis in the present paper, but for the most part we will
ignore it.

The interactions of localizable states are well approximated by
quantum field theory as long as no black holes are formed in
scattering.   For black holes that are much larger than the Planck
scale but much smaller than the horizon radius, the inclusive
formulae of Hawking evaporation give a reasonably accurate
description and it is unrealistic to hope for a practical
calculation which could give more accurate information than these.

The entropy of localized states obeying these constraints is of
order $(RM_P)^{3/2}$.  Consequently, there is a limit to the
accuracy of measurements that can in principle be made of $S_R$. It
is of order $e^{ - c (RM_P)^{3/2}}$, with $c \sim 1$.   Another way
of saying this is that the pointers of any conceivable measuring
apparatus built from these degrees of freedom, will undergo
spontaneous quantum jumps between pointer positions, on time scales
of order $T_q \equiv e^{  c (RM_P)^{3/2}}$ (units are essentially
irrelevant for these huge numbers when $RM_P$ is a few orders of
magnitude or greater).   This time scale is much shorter than the
{\it recurrence time}\cite{dyson}.   More relevant for our
discussion will be that in many cases it is much shorter than the
tunneling times between false and true vacua.

We should also note that the probability of an observer making
observations on time scales of order $T_q$ is itself extremely
small. For example the probability of spontaneous nucleation, at the
observer's position, of a black hole of mass $M$ is just $e^{- 2\pi
RM }$.  For large $RM_P$ the probability of an observer surviving
for a time $T_q$ is superexponentially small.

It will be important in the sequel that {\it most} of the states of
the dS quantum theory decouple from the system when we take the
$\Lambda \rightarrow 0$ limit.   The limiting theory is described by
an S matrix, which takes into account only localizable states in a
single dS horizon, whose Poincare energy remains finite in the
$\Lambda \rightarrow 0$ limit.

\subsection{dS transitions}\label{dstransitions}

The CDL instanton for dS to dS transitions is a compact Euclidean
manifold with the form of an ovoid:
\begin{equation}
ds^2 = dz^2 + \rho^2 (z) d\Omega^2 ,
\end{equation}
accompanied by a scalar field configuration $\phi (z)$.   The $z$
coordinate lives in the interval $[0, L]$ and $\rho (z)$ and
$\dot{\phi} (z)$ vanish at both ends of the interval.   The
instanton exists unless the maximum of the potential is too
flat\footnote{The interpretation of this is that tunneling occurs
for a flat maximum, but the semi-classical approximation breaks
down.  The system tunnels from a region near one minimum of the
potential to a position on the flat maximum, where quantum
fluctuations are large, and then rolls off the other end.}.  In
general, there will be a finite\footnote{In \cite{heretics} one of
the authors (T.B.) made the incorrect claim that there were an
infinite number of solutions and that at least one always existed.
Many colleagues, including A.Linde, E.Weinberg, and most recently
M.Kleban, set him straight on this.} discrete set of such instanton
configurations, but only one where the scalar field makes only a
single pass over the maximum separating the two minima, as $z$
ranges over the interval. The field never hits either the false or
the true vacuum point. The fact that it does not hit the false
vacuum is interpreted physically as the statement that we are
describing a thermal system. The instanton represents the most
advantageous thermally activated process for jumping over the
barrier, rather than simple vacuum tunneling. Mathematically, the
same fact is explained by noting that the instanton is compact. It
is asymptotic boundary conditions on an infinite $z$ interval, which
force the instantons for decay of an asymptotically flat or AdS
space to asymptote to the false vacuum.

The Lorentzian continuation of the instanton describes two causally
disconnected ``bubbles" separated by a static (in appropriate
coordinates) wall region, which is typically of microscopic size
even if parameters have been tuned to make both dS radii large. Each
bubble is an excitation of a de Sitter background, one in the true,
and one in the false vacuum.   Causal diamonds of observers in these
bubbles have finite maximal area, and inside a causal diamond the
system quickly reverts to the dS vacuum configuration.

The action of the instanton, $S_I$ is negative and it is converted
into a probability by subtraction of the dS action of either the
true or false vacua.  We thereby compute two transition
probabilities
\begin{equation}
P_{T \rightarrow F} = e^{- (S_I - S_T)},
\end{equation}
and
\begin{equation}
P_{F \rightarrow T} = e^{ - (S_I - S_F)}.
\end{equation}
They are related by
\begin{equation}
P_{T \rightarrow F} = e^{- (S_F - S_T)}P_{F \rightarrow T}.
\end{equation}
We believe that the paper\cite{lw} is probably the first to estimate
the inverse transition amplitude from a true dS vacuum to a false
one. Recognizing that the absolute value of the Euclidean dS action
is equal to dS entropy we find that
\begin{equation}
P_{T \rightarrow F} = e^{- (\Delta {\rm Entropy})}P_{F \rightarrow
T}.
\end{equation}
This is the degenerate form of the law of detailed balance for
systems whose states are all at energies much lower than the
temperature.  Thus, the CDL analysis is not only consistent with our
picture of dS space as a system with a finite number of states.  It
even confirms the conclusion\cite{tbds} that the bulk of the states
lie at energies below the dS temperature.

There is a class of solutions of the CDL equations, for which the
intuition that gravity does not make large contributions to the
tunneling amplitude is correct.  Usually this is attributed to the
fact that the true vacuum bubble size, upon nucleation is much
smaller than the dS horizon. We will try to argue that this
intuition is incorrect. It is convenient to work with a class of
potentials of the form $\mu^4 v(\phi / M)$, where $\mu \ll m_P$ and
$\mu \ll M \leq m_P$. $v$ is a bounded function of its argument, $x
\equiv \phi /M$. Very often, we work with polynomial potentials,
motivated by the renormalizability criterion of low energy effective
field theory. This is perfectly permissible as long as the processes
we study do not force us to explore asymptotically large values of
$\phi/M$. The condition $\mu \ll M$ is the criterion for validity of
the semi-classical approximation:  all tunneling probabilities will
be very small in this limit.   For potentials of this form, and in
the absence of gravity, the length scale of variation of the
instanton is ${M\over {\mu^2}}$.  We will work in dimensionless
units, with this parameter being the standard of all length and
energy scales. The non-gravitational instanton action is of order
$(M/\mu)^4 $. The parameter $\epsilon^2  \equiv {{M^2}\over {3
m_P^2}}$ measures the size of the gravitational corrections to the
tunneling equations, but we will see that the corrections are
singular perturbations.

The metric and scalar field for the instanton have the form
\begin{equation}
ds^2 = dz^2 + \rho^2 (z) d\Omega^2,
\end{equation}
\begin{equation}
\phi = \phi (z) .
\end{equation}
The metric is that of a lozenge or ovoid, because for $dS$ tunneling
the $z$ interval is compact, $z \in [0, z_f ]$.  Define
\begin{equation}
x \equiv \phi / M ,
\end{equation}
\begin{equation}
r \equiv {\mu^2 \rho \over M},
\end{equation}
\begin{equation}
s \equiv {\mu^2 z \over M},
\end{equation}
and
\begin{equation}
\epsilon^2 = {M^2 \over 3 m_P^2}.
\end{equation}
Then the CDL equations are
\begin{equation}\label{dotr}
\dot{r}^2 = 1 + \epsilon^2 r^2 ({{\dot{x}^2}\over 2} - v(x)),
\end{equation}
\begin{equation}\label{ddotr}
\ddot{r} = -\epsilon^2 r \left( \dot{x}^2 + v(x) \right),
\end{equation}
\begin{equation}\label{ddotx}
\ddot{x} + 3{{\dot{r}}\over r} \dot{x} = v^{\prime} (x).
\end{equation}
Here dots refer to derivatives with respect to $s$, and
Eq.~\ref{ddotr} is found from Eq.~\ref{dotr} and \ref{ddotx}. We
define
\begin{equation}
E \equiv {1\over 2} \dot{x}^2 - v ,
\end{equation}
which satisfies
\begin{equation}
\dot{E} = - 3 {{\dot{r}\over r}}(\dot{x})^2 .
\end{equation}
Note that this Euclidean energy decreases when $r$ increases, and
vice versa.

The boundary conditions are that we have a smooth, compact manifold,
corresponding to
\begin{equation}
r(0) = r(s_f) = 0 =\dot{x} (0) = \dot{x} (s_f).
\end{equation}
$s_f$ is freely chosen to satisfy these conditions. However, the it
is more convenient to parametrize the unique boundary condition in
terms of $x(0)$, and let $s_f$ be determined by the equation $r(s_f)
= 0$. Later we will write a form of the equations in which $s$ is
eliminated in favor of $r$.

Note that it is the equation for $\rho$, which requires these
boundary conditions, which, when the false vacuum has positive
energy density, are infinitely different from those for
non-gravitational instantons. The small parameter $\epsilon$ is a
singular perturbation of the non-gravitational equations.  For any
non-zero value of $\epsilon$, and any non-singular solution of the
equations, $\rho$ increases and then decreases back to zero.  The
easiest way to see this is to note that the equations are identical
to the equations for a Lorentzian cosmology with negative
cosmological constant, whose generic solution ends in a Big
Crunch\footnote{But please do not confuse this Big Crunch analogy,
with the genuine Big Crunch cosmology which we find in the
Lorentzian continuation of the instanton when $v_T < 0$.}. Here we
can avoid the Crunch by exploiting the freedom to choose the value
of $\phi$ at (either one but not both of) the end points. As a
consequence, even when one of the dS minima has very small c.c., the
solution does not begin or end exactly at the minimum of the
potential. Loosely speaking, this is a reflection of the thermal
nature of dS space. The CDL instanton reflects both vacuum tunneling
and thermal activation.   Indeed, given the picture of the static dS
Hamiltonian advocated in \cite{tbds}, the two are the same.   The dS
vacuum is in fact a thermal ensemble of a very dense set of levels,
clustered below the dS temperature.

When $\epsilon \ll 1$, we see a problem in solving these equations.
The interval of $s$ over which the radius of the ovoid expands and
recontracts, is of order $1\over \epsilon$, while the natural
(Euclidean) time scale of variation of $x$ is of order $1$.  Thus,
if we are looking for an instanton solution which does not have
oscillations of $x$\footnote{The oscillating instantons describe
transitions with smaller probability than the one which makes a
single pass over the maximum of $v$ at $x_H$.}, we must find a place
where $x$ can sit for an ``unnaturally" long time. The only such
places are the stationary points of the potential. We will assume
that the curvature of $v$ is of order $1$ at all these points.   A
solution which starts very near the maximum of $v$ at $x_H$ will
oscillate around the maximum with frequency of order one, and does
not describe a single pass instanton.

Suppose we start with a solution which starts\footnote{We can
``start" the instanton in the basin of attraction of either $x_T$ or
$x_F$. These are two different descriptions of the same solution.}
with $x(0) \sim x_F$. We will refer to Figure 2, where $x_F$ is the
leftmost point on the potential, which might be visited by the
instanton, and choose $s =0$ to be the coordinates of the leftmost
point in the instanton. For $s$ near zero, we can solve the
equations exactly, and find that
\begin{equation}\label{bessel}
x(s) \approx x_F + (x - x_F){{I_1 (\omega_F t)} \over{\omega_F t}},
\end{equation}
where $I_1$ is the Bessel function of imaginary argument, and
$\omega_F$ the curvature of $v$ near $x_F$. The solution increases
exponentially at large $\omega t$, where the approximation breaks
down.   However, we can trust it until the deviation is of order
$1$.  At this point we are a distance of order $1$ from $x_T$, and
have a velocity of order $1$. Recall that the CDL equations describe
a particle moving in the inverse potential $- v$. The system begins
with friction in the $x$ equation. Anti-friction, which could lead
to a singularity, sets in only after a time of order $1/\epsilon$
when $r$ starts to re-contract. Thus, in general, the solution
oscillates around $x_H$, the Hawking-Moss maximum of $v$ (see Figure
2.). If we want a single pass instanton, we must, for small
$\epsilon$ choose $(x - x_F)$ exponentially small, in order that we
do not run away from the minimum before $r$ expands to its maximum,
which is $r_m = o({1\over \epsilon })$, and recontracts to $r =
o(1)$. \EPSFIGURE{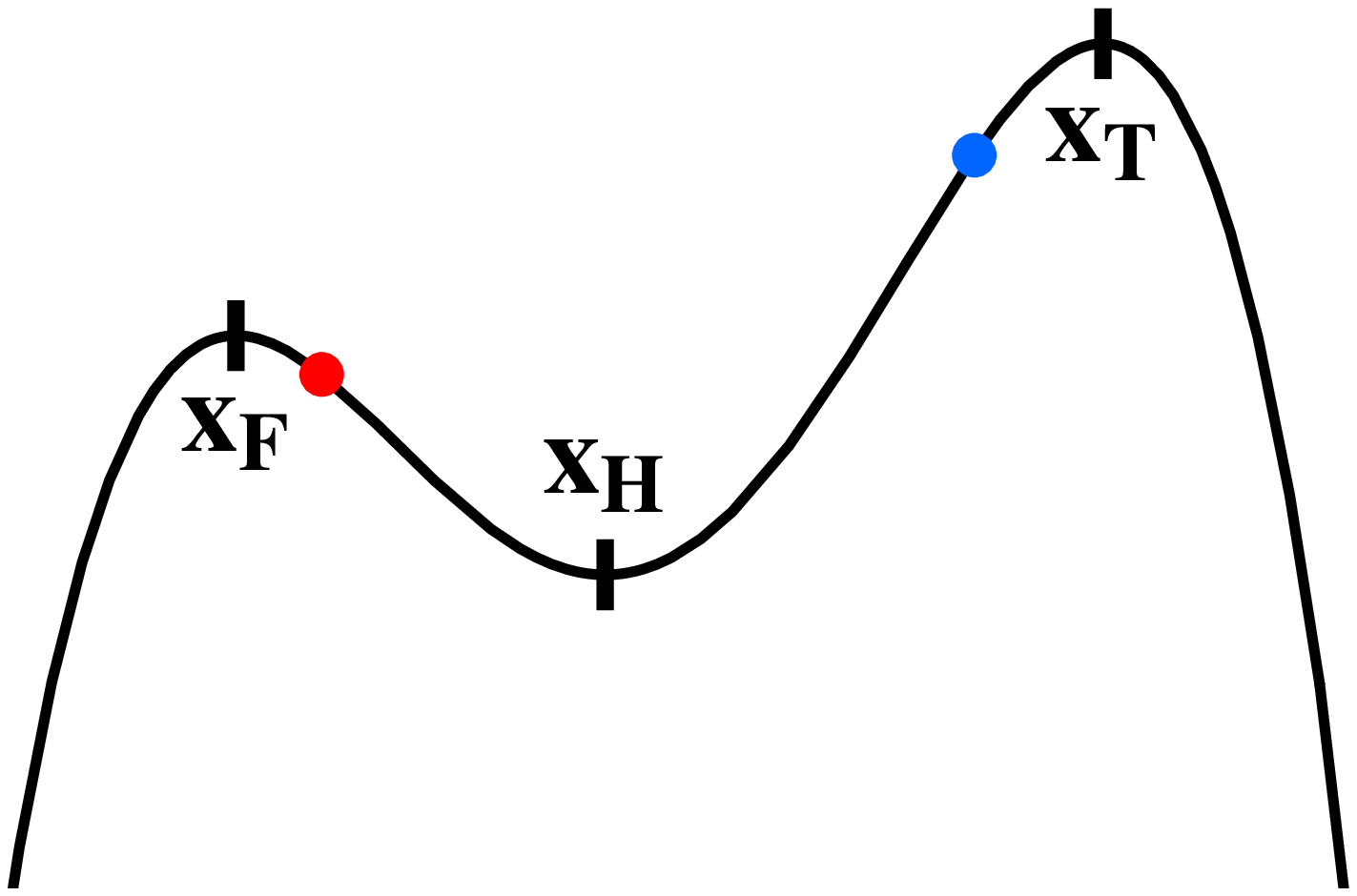,width=75mm}{The Instanton Moves In The
Inverted Potential $-v$. A Set of Sample Turning Points are
Indicated (Red and Blue Dots, Corresponding to the Initial or Final
Positions). The Words Left and Right in the Text Refer to This
Figure.\label{uofx}}

The Euclidean energy
\begin{equation}
E = {{{\dot{x}^2}}\over 2} - v(x),
\end{equation}
starts at $- V_F$ and stays close to this value until $r$ hits
$r_m$. Then it begins to increase.   If we had instead started very
close to $x_T$ (so that the energy was close to $- V_T$) this would
have implied a disaster. A non-singular instanton must stop at a
point on the potential to the right of $x_F$ in Figure 2, but a
solution stopped at that point has energy $ < - V_T$.  Initial
conditions very close to $x_T$ give singular solutions.

However, starting as we have, near $x_F$, there is no problem
(assuming the barrier is not too flat) in finding a non-singular
solution. Note that although the criteria for the thin wall
approximation are not generally satisfied, it remains true, for $
\epsilon \ll 1$, that the instanton resembles the false vacuum over
most of its volume. Since the solution spends most of its time near
$x_F$, the action difference $S_I - S_{dS_F}$ is positive and
independent of $\epsilon$ and we get a transition probability of
order $e^{ - c(M/\mu)^4}$.  The exponent is of the order of
magnitude we would have expected in a calculation which neglected
gravitation.  The probability for the inverse transition is, on the
other hand, superexponentially suppressed when the true c.c. is
exponentially small.

As $\epsilon$ is raised to $1$, the initial value $x(0)$ moves
further from $x_F$, corresponding to the increasing Gibbons-Hawking
temperature.  The instanton action is of order $c (m_P /\mu)^4$ with
a constant $c$ of order $- 1$.   The probability of true to false
transitions is of order $e^{- d {{m_P^4}\over V_T}}$, with a
constant $d$ of order $1$.  {\it Now suppose we change the potential
only in the region closer to $x_F$ than $x(0)$, in such a way that
the false dS minimum moves down to negative energy.   The instanton
we have found is still a solution of the equations for this new
potential, and predicts a probability for transition from the small
c.c. dS space to a negative energy Big Crunch, which is similarly
exponentially suppressed.}  This is our first indication that the
conventional picture of dS to Big Crunch transitions, is flawed.  Of
course, an advocate of the conventional picture could still contend
that there is another instanton, which mediates the transition with
higher probability.   We will postpone till the next section the
rebuttal to this argument.

As a consequence of the law of detailed balance obeyed by $dS
\rightarrow dS$ transitions, we proposed in
\cite{heretics}\cite{tbds} a regulated description of eternal
inflation.   The conventional picture views the false vacuum as
dominating the global structure of the universe. It expands much
more rapidly than it decays and so volume weighted averages over
appropriately chosen spatial slices are dominated by the false
vacuum.  Inflation is viewed as occurring eternally, ``somewhere in
the multiverse".  Our picture is very different.  The false vacuum
is a finite, low entropy subsystem of the dominant true vacuum
region, which itself has a finite number of physical states. In the
false vacuum ensemble, most of the degrees of freedom of the
universe are frozen into a very special state. This configuration
decays relatively rapidly, back to the true vacuum

The true vacuum on the other hand is essentially stable.  For
example, once $(R_T M_P)^2 > 10^{23}$ it is no more likely for the
true vacuum to jump to the false one than it is for the proverbial
gas in a box to spontaneously collect in a corner. Indeed, for large
values of $R_T M_P$ it is not clear that there is any operational
meaning to these jumps. If we view an observer as a naked world
line, then it is true that it will encounter an infinite number of
back and forth jumps in the course of its history in dS space. On
the other hand, if we insist that an observer be a large localizable
system with robust semi-classical observables, then the typical
observer will probably be blasted apart by spontaneously produced
thermal radiation, swallowed by spontaneously nucleated black holes,
or suffer quantum jumps of its pointer variables, an exponentially
large number of times before it has to endure a CDL tunneling event.

To conclude, it is our contention that the wildly fluctuating,
eternally inflating (at the false Hubble parameter), universe of the
conventional description, is a figment of the quantum field
theorist's imagination.   The true theory of quantum gravity will
view the dS to dS transitions as a more or less conventional thermal
description of transitions in a finite system, to and from a low
entropy meta-stable state.  Note that even within the conventional
description, no actual experiment performed by local observers will
behave differently than our alternate picture predicts.   It is only
when we take the God-like view of the field theorist and try to
interpret global properties of a multiverse that can never be
observed, that we run into confusion.

\section{The crunch at the end of the tunnel}\label{crunch}

We now want to generalize these considerations to the case where the
true vacuum has negative c.c.   As shown by CDL, the Lorentzian
bubble in this case, undergoes a Big Crunch on a time scale
determined by the shape of the potential in the region of negative
energy density. From the eternal inflation point of view, this is
not necessarily a problem. In tunneling from dS space, there is
always a second bubble which remains in the dS phase.  In the
conventional view of eternal inflation, any given observer, with
probability one, eventually finds himself in the Big Crunch, but
``the overwhelming volume of the multiverse is eternally inflating".
From this point of view, the ratio $r = |V_T|/|V_F|$ does not change
the qualitative nature of the physics.

The holographic interpretation of these transitions is quite
different, and the ratio $r$ plays a crucial role.  We will begin
with the regime of large $r$, which is easiest to interpret. We will
argue that transition probability from false to true vacuum is, when
$RM_P$ goes to infinity, of order $e^{ - c (RM_P)^2}$, where $R$ is
the radius of the false dS vacuum. A crucial observation is that
within the true vacuum bubble of the CDL instanton, there is a
maximal area causal diamond, and the maximal area is of order
${m_P^4 \over {| V_T|}}$. In making this estimate, we have imagined
the same form for the potential as in previous sections: $\mu^4
v(\phi / M)$, where $\mu$ and $M$ are two mass scales with $\mu \ll
m_P$ and $M \leq m_P$. $v$ is a function with two non-degenerate
local minima. We will later adjust dimensionless parameters in the
potential to tune the higher minimum to a very small positive value,
while the negative minimum remains at a microphysical scale.

The action of the instanton is negative, because the lozenge has
positive curvature, and the Einstein term dominates the matter
action as a consequence of the gravitational field equations.
Indeed, the Euclidean Einstein equations give
\begin{equation}
m_P^2 (R_{\mu\nu} - {1\over 2} g_{\mu\nu} R) = \partial_{\mu}\phi
\partial_{\nu} \phi - {{g_{\mu\nu}}\over 2} ([\nabla\phi ]^2 +
V(\phi )),
\end{equation}
which implies
\begin{equation}
m_P^2 R = [\nabla\phi ]^2 + 2 V,
\end{equation}
and
\begin{equation}\label{action}
S = - \int \sqrt{g} V = - 2\pi^2 ({M\over\mu})^4 \int_0^{s_f} [r^3
v],
\end{equation}
for a potential of the form we have described.  We will see that the
potential is positive over much of the volume of the instanton, so
that the action is negative.

It is worth understanding how the solution that describes tunneling
of an asymptotically flat space into a negative c.c. region, emerges
in the limit of vanishing c.c. .   This solution is described by the
same differential equations.   However, if $v(x_F) = 0$, we see the
possibility of a solution where $z$ is non-compact and $r$
asymptotes to infinity like $r = z$.   All that is necessary is that
$E$ vanishes sufficiently rapidly that $r^2 E \rightarrow 0$.  In
fact, examining the equations in the large r regime, we see that $E
\sim e^{- b r} $ is the actual behavior.   $b$ is proportional to
the square root of the curvature of the potential at the false
vacuum.   For such solutions, the gravitational correction
$\epsilon^2 E r^2$ is a uniformly small perturbation.   The solution
is well approximated by its non-gravitational counterpart. Its
action is positive and of order $({M\over \mu})^4$, with small
corrections of order $\epsilon^2$.   This answer is very different
both in sign and magnitude from the action for {\it any} compact
instanton.  However, we have to recall the subtraction of the action
of the false vacuum dS space. An instanton which loiters near the
false vacuum, so that it is equal to the false dS space over most of
its volume, could have a positive action difference of order
$({M\over\mu} )^4$.

{\it Thus, the question we now want to answer is whether, for very
small but non-vanishing $v(x_F)$ the instanton converges to the flat
space answer.  We claimed above that the answer is NO.} To see this,
begin at $s = 0$ and some point $x(0) = x_I$, in the basin of
attraction of the true vacuum\footnote{For this paragraph, and this
paragraph only, the Euclidean time runs from the right to the left
of the ovoid, starting near $x_T$ and ending near $x_F$ in Figure
2.}. Both $\dot{r} $ and $\dot{x}$ are of order one. In general,
unless we approach a point where $E$ is very small and $\dot{x}$ can
remain small (the only such point is $x_F$), the point at which the
expansion of $r$ turns around will occur at $s \sim
{1\over\epsilon}$ and $r \sim {1\over\epsilon}$, independent of
$v(x_F)$.   For such solutions, the maximal radius is of
microphysical size, and there is no resemblance to the
non-gravitational instanton. A solution with $r_{max} \sim {1\over
\epsilon\sqrt{v(x_F)}}$ can only be obtained by tuning $x_I$ so that
one approaches $x_F$ from the right (right and left refer to Figure
2.) with very small velocity, while the radius is still expanding.
Friction and the flatness of the potential in the vicinity of $x_F$
will then guarantee that $r$ increases until $\epsilon^2 r^2 v(x_F)
\sim 1$. So the turnaround occurs with $x \sim x_F$ and $0 > \dot{x}
$ and $|\dot{x}| \ll 1$. But now we may be lost.   $r$ begins to
decrease, slowly at first, and then at an $o(1)$ rate, so we must
wait a long time until $r = 0$ again.   $x$ on the other hand, is
subject to ever increasing anti-friction.  We claim that this will
overwhelm the tiny force near the top of the barrier in $- v$. The
field will sail past $x_F$ and the solution will be singular.

To be even more explicit, let us note that we can parametrize our
solutions by $r_{max}$, the maximal radius of a cross section of the
ovoid, and the value $x_{max}$ at which this maximum is achieved.
The relation
\begin{equation}
\epsilon^2 r_{max}^2 = - {1\over E_{max}} = - {1\over {{(\dot{x})^2
\over 2} - v(x_{max})}},
\end{equation}
determines the velocity at this point and thus completely determines
the solution of the equations.   We prefer to start at this point
and consider taking the radius to zero separately on the left and
right halves of the ovoid. For a single pass instanton, $x$
increases on the right half, and decreases on the left half of the
ovoid, as $r$ decreases to zero.

On each branch, we can write the equations for the trajectory
$x^{\pm} (r)$ ($+$ refers to the left, and $-$ to the right half of
the ovoid), as $r$ varies between $r_{max} $ and $0$. They are
\begin{equation}
E_r = {- {6\over r}} (E + v(x)),
\end{equation}
\begin{equation}
x^{\pm}_r = \pm \sqrt{{2( E + v(x))}\over {1 + \epsilon^2 E r^2}}.
\end{equation}
We also have the constraint $E + v(x) \geq 0$.  We can solve the
first equation to obtain
\begin{equation}
E^{\pm}  = E_{max} ({{r_{max}}\over r})^{- 6} + 6 \int_r^{r_{max}}\
dy\ [y^5 v( x^{\pm}(y))] r^{-6} .
\end{equation}
This is singular at $r = 0$ unless
\begin{equation}
E_{max} = - 6 r_{max}^{- 6} \int_0^{r_{max}}\ dy\ [y^5 v
(x_{\pm}(y))].
\end{equation}
We can then rewrite
\begin{equation}
E^{\pm} (r) = - 6 \int_0^r dy [ y^5 v(x^{\pm} (y))] r^{-6}.
\end{equation}
In order to have a geometry which approaches the flat space
instanton, we must have $\epsilon^2 r_{max}^2 = - {1\over E_{max}}
\sim  {1\over {v(x_F)}} \gg 1.$  On the other hand, our formula
gives
$$E_{max} = o(1),$$ unless the instanton loiters near $x_F$
for most of its trajectory.   Thus, for large dS radius, this
solution must have $x_{max} \approx x_F$, and of course $x_F <
x_{max} < x_T$ .  We can certainly find a smooth solution on the
left half of the ovoid which has this value of $x_{max}$, using the
same undershoot/overshoot argument that works for the flat space
case\cite{coleman}.

However, now we must complete the right half of the ovoid.   The
equations tell us that near $r_{max}$, ${dx\over dr}$ is large and
positive.   Thus, as $r$ is lowered on the right side of the ovoid,
$x$ quickly passes to the right of $x_H$.   After that it is
downhill all the way.   From the point of view of the cosmological
time $s$, the ``universe" is contracting.   The resulting
anti-frictional force pulls in the same direction as the potential,
and we reach infinite negative $E$ in finite time.   The curvature
of the metric is singular at the same point.

We conclude that no regular geometry can approach the flat space
solution.  Instead, that solution is gotten by taking the limit of a
singular solution of the finite radius equations.  The singular half
of the large ``compact" singular space is thrown away as we take
$V(x_F) \rightarrow 0$, and we obtain a regular solution with
asymptotically Euclidean boundary conditions.

These arguments sound convincing, but are not rigorous.   Things are
particularly confusing for $\epsilon \ll 1$.   In this case, since
it takes time of order $1/\epsilon$ for $r$ to expand to its maximum
and recontract, there are lots of non-singular solutions which
oscillate of order $N \sim {1\over \epsilon}$ times around the
maximum of $v$ at $x_H$.   If $N$ is odd these can be interpreted as
CDL tunneling to a true and false vacuum bubble, but there will also
be a solution which makes only a single traverse over the maximum.
This solution loiters around the false vacuum for a time at least as
long as $1\over \epsilon$, while $r$ expands to its maximum (also at
least of order $1\over \epsilon$) and recontracts to a value of
order $1$. It then crosses over to the basin of attraction of $x_T$
in a time of order $1$.    The question is, if $\epsilon$ is held
fixed and parameters tuned so that $V_F$ goes to zero, does the
loitering time go like ${1\over \epsilon\sqrt{V_F}}$, so that the
solution looks like the false dS space over most of its volume?  Or
does the loitering time asymptote to a constant as $V_F$ goes to
zero (and how does the constant depend on $\epsilon$)?

In order to loiter near the false vacuum for a time of order
${1\over \epsilon\sqrt{V_F}}$, the value of $x$ at the leftmost zero
of $r$ must be exponentially close to $x_F$.  Indeed, in this regime
we can solve the equations exactly and
\begin{equation}
x(s) = (x(0)- x_F){{I_1 (\omega_F s)}\over {\omega_F s}},
\end{equation}
where $I_1$ is the Bessel function of imaginary argument, $\omega_F$
is the curvature of $v$ at $x_F$.

For application to models of low energy SUSY breaking, we want to
tune the false minimum to zero via a formula like
\begin{equation}
v_z (x) \equiv f(x) - (1 + z) f(x_F) .
\end{equation}
We claim that as $z \downarrow 0$ $x(0) - x_F$ converges to a
non-zero, $\epsilon$ dependent, constant. Unfortunately, we have not
been able to come up with a rigorous analytic argument that this is
so for this case. Instead, we have resorted to numerical
calculations.   Results of a preliminary study are reported in the
next section.  We hope to return to a more detailed numerical
investigation of this question in a forthcoming paper\cite{abj}

 The action for
the smooth compact instanton is negative and of order $({m_P \over
\mu})^4 $. Thus, as in the case of the $dS \rightarrow dS$
transitions, we must make a subtraction in order to define a
probability.   For the transition from ``false" to ``true" vacuum,
it is obvious that the correct formula should be
\begin{equation}\label{Pft}
P_{F\rightarrow T} \sim e^{- (S_I - S_{dS})} .
\end{equation}
If we are correct that $S_I$ asymptotes to a constant for large de
Sitter radius, then this is an incredibly tiny number, essentially
the inverse of the recurrence time for the large de Sitter space.

We have gone to great lengths to {\it explain} the mathematical
behavior of the CDL amplitude for the case $|V_T| \gg V_F > 0$,
because we have found so many people who are surprised by the
answer. It is worth pointing out however that the result is not new.
It was mentioned explicitly in \cite{kklt} and is at least implicit
in many much earlier papers.  In particular the paper
\cite{lindeopen} contains similar estimates and discusses their
relation to the entropy of dS space.  As far as we know however,
there has been no previous attempt to explain the physical basis for
the mathematical discontinuity of the result when $V_F = 0$.

There {\it is} a simple physical explanation, based on the picture
of dS quantum mechanics proposed in \cite{tbds}, of the peculiar
mathematical discontinuity of the CDL instanton as the cosmological
constant in the ``false" vacuum is taken to zero.   The flat space
limit, in this picture, is obtained by ignoring most of the states
of dS space and concentrating on the low energy localizable states,
classified as eigenstates of the operator $P_0$.   For finite dS
radius, the localized states are unstable, and decay back to the dS
vacuum ensemble, but as $R \rightarrow\infty$ the lifetime of
localizable states becomes infinity and $P_0$ becomes a conserved
quantity.   The CDL instanton for asymptotically flat space decaying
to a Big Crunch describes the properties of the unique zero energy
eigenstate of $P_0$.

By contrast, for finite $R$ the vacuum of the theory is a thermal
ensemble with (for large $R$) huge entropy.   Our mathematical
results suggest that rather than a decay, the CDL instanton for this
case describes an unlikely fluctuation to a low entropy meta-stable
state.   We can find further evidence for this interpretation by
applying the covariant entropy bound to the Big Crunch portion of
the Lorentzian CDL instanton.   For the potential we have chosen, it
is easy to see that the area in Planck units of the holographic
screen of the maximal causal diamond for observers in this portion
of the space-time is of order $({m_P \over\mu})^4$.   To see this
use a dimensionless time variable, field, and scale factor, with the
same scaling factors we used in the Euclidean calculation above. The
dimensionless cosmological time to the crunch is of order
${1\over\epsilon}$, as is the value of the dimensionless scale
factor at that time.  The actual proper time and scale factor are
both of order ${{m_P} \over {\mu^2}}$.   The coordinate distance to
the particle horizon scales like:
\begin{equation}
d_H \sim \int {dt\over a(t)} \sim 1.
\end{equation}
The area of the holographic screen of the ``last observer standing"
is thus $\sim m_P^2 a^2 \sim {{m_P} \over {\mu^2}}$.  Thus, there is
reason to believe that the Big Crunch is a low entropy system,
compared to the dS vacuum ensemble, when $\mu$ is a microphysical
scale but the ``false" vacuum energy is tuned to be near zero.

We would thus advocate a calculation of the probability of the
reverse tunneling event: Big Crunch to dS of the form,
\begin{equation}
P_{T \rightarrow F} \sim e^{-(S_I + {1\over 4} A_{BC})},
\end{equation}
where $A_{BC}$ is the area, in Planck units of the largest causal
diamond in the Crunch region of the CDL instanton.   This formula,
together with the one we have given for the inverse transition,
would satisfy the law of detailed balance if we assumed that the
entropy dominates the free energy for both the dS space and Big
Crunch.   In the case of dS we have already explained that this is
consistent with the spectrum of the dS Hamiltonian advocated in
\cite{tbds}.   For the Big Crunch space-time there is no conserved
energy, and it is not clear what we mean by free energy. However, it
is reasonable to assume that the system there maximizes its entropy,
without any constraints\cite{blacrunch}. The rapid time dependence
of the fields will excite all degrees of freedom available to the
system.

If this interpretation of the CDL instanton is correct, we again see
a complete disagreement with the logic of eternal inflation.
Consider observers in the dS and Crunch regions of the CDL diagram,
assuming the covariant entropy bound. The {\it
heretics}\cite{heretics} which remain in the false vacuum can access
a huge amount of entropy and exist for an extremely long time.  The
{\it true believers} by contrast, exist for a microscopically short
period and can access much less information.

The conventional phrases ``false and true vacua", and
``instability", seem misleading in this context.   The true vacuum
is not a stable state in any sense.   Inhomogeneous and anisotropic
classical perturbations grow as we approach the Crunch. If we tried
to do quantum field theory in the interior of the CDL bubble, using
a Gaussian wave functional describing small fluctuations around the
bubble as the initial ``vacuum state" and polynomials in field
operators acting on this state as the excitations, then the state
near the Crunch is full of excitations.    Everything seems to
indicate that the system wants to ``thermalize" in the sense of
exploring the entire Hilbert space of states available to it. The
covariant Entropy bound suggests that this space of states is finite
dimensional, with relatively small dimension.   On the other hand,
the smaller the value of the positive c.c., the more the
``meta-stable" false vacuum looks like a stable vacuum state.  It
has a temperature, which goes to zero with the c.c.   This means
that excitations occur, but with very small probability.

As small as it is, the probability for even outlandish events like
nucleation of a galaxy size black hole, is much more probable for
small c.c. (dS radius much larger than a galaxy) than the CDL
transition to the ``true vacuum". This is easy to understand: the
thermal probability for black hole production, $e^{- m/T}$ just
reflects the entropy deficit(!) of black hole states relative to
states of the empty dS vacuum\cite{tbds}. The entropy of the ``true
vacuum" is much smaller than that of large black holes.

When the absolute value of the negative c.c. is much smaller than
the positive false vacuum energy, the maximal causal diamond in the
Big Crunch space-time is much larger than the dS horizon size.   In
this case, it is plausible that the situation is reversed and that
the "true vacuum" really is the state in which the system spends
most of its time.  However, we are much less certain that there is a
real model of quantum gravity corresponding to this semi-classical
picture.   Indeed, quite generally, when we try to understand the
quantum meaning of all possible solutions of general relativity
coupled to other matter fields, we may be trying to make sense out
of utter nonsense.   String theory has taught us that theories of
quantum gravity are much more restrictive than classical field
theory would lead us to believe.  We have no solid evidence that a
real theory of quantum gravity will produce {\it any} of the
situations studied in this paper.  It should not surprise us if some
of them resist interpretation.

In both the situation where $- V_T \gg V_F$ and $ - V_T \ll V_F$ the
model contains no immortal observers.  However, Observer's Doomsday
is much more immediate in the latter case.   The time before the Big
Crunch is power law in $|V_T|$.   By contrast, processes which
destroy typical observers in the case where the ``false" dS vacuum
is the ground state (according to our criteria) have probabilities
that are superexponentially suppressed as the c.c. goes to zero.

Finally we would like to note that, although we have give a
satisfactory explanation of the discontinuity in instanton
amplitudes as the dS radius goes to infinity, it might be that this
is a fact that does not need an explanation.   If the conjecture of
\cite{tbfolly} is correct and the universe necessarily becomes
supersymmetric as the c.c. is sent to zero, then in real quantum
theories of gravity, Poincare invariant vacuum tunneling to a Big
Crunch never occurs.

\section{Some numerical results}

In the previous section, we put forward a set of arguments which
imply that there is no regular instanton which as we take $V(x_{F})
\rightarrow 0$ approaches the flat space  solution. We conjectured
that there {\em is} a regular instanton which interpolates between
the wells of the potential, but in the limit where $V(x_{F})
\rightarrow 0$, this solution will have $r_{max} \sim
\frac{1}{\epsilon}$, independent of the value of $V(x_{F})$.
Unfortunately, our analytics fall short of a proof, and so in this
section we demonstrate some examples of this behavior using
numerical methods.

The model we use for the potential is:
\begin{equation}\label{potential}
v_{z} \left(x\right) =f \left( x \right) -\left(1+z\right) f \left(
x_{F} \right),
\end{equation}
with
\begin{equation}
f(x) =\frac{1}{4} x^4 - \frac{b}{3} x^3 -\frac{1}{2} x^2.
\end{equation}
This auxiliary function, $f(x)$, has a local maximum at the origin,
and two negative minima. The potential $v_{z}(x)$ for three
representative values of the parameter $b$, which controls the
relative height of the two minima, is shown in
Fig.~\ref{potentials}. Also shown in this figure (see the inset), is
an example of the behavior of a potential (with $b=.3$) as $z$ is
taken to zero. In this limit, the false vacuum is a de Sitter space
with vanishing cosmological constant.

\EPSFIGURE{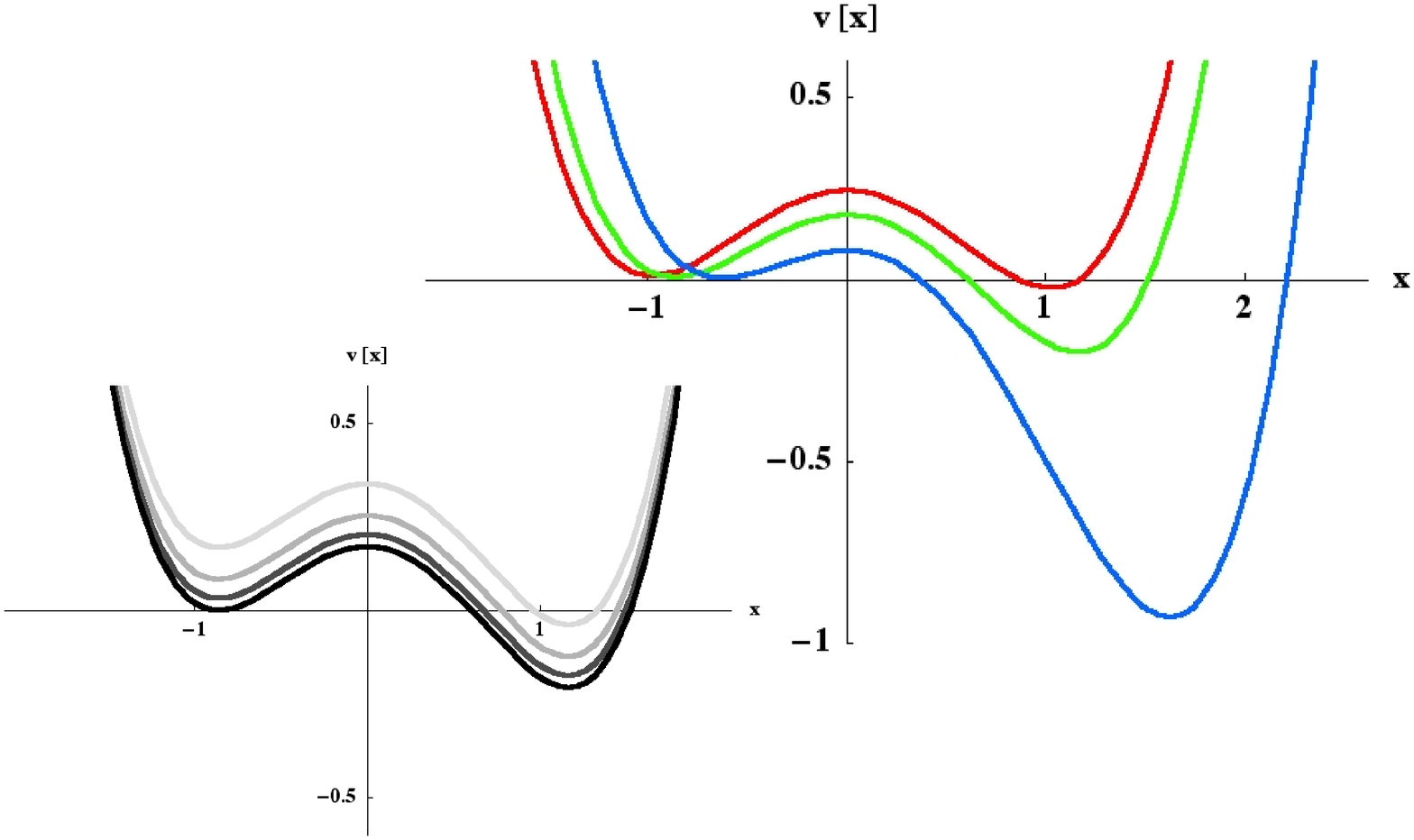,width=150mm}{The potential $v_{z} (x)$ for
$b =\{.05,~.3,~1\}$ (red, green, blue) with $z=.05$. It can be seen
that the parameter $b$ controls the relative heights of the true and
false vacuum wells. The inset shows the potential with $b=.3$ and
$z=\{1,~.5,~.2,~.01\}$ (light to dark). As $z \rightarrow 0$, the
false vacuum approaches a dS space with vanishing cosmological
constant. \label{potentials}}

The equations of motion in this potential are given by
Eq.~\ref{dotr}, \ref{ddotr}, and \ref{ddotx}. We will use
Eq.~\ref{ddotr} and~\ref{ddotx} to evolve, with the initial
conditions $x\left(0 \right) = x_{I}$,  $\dot{x} \left( 0
\right)=0$, $r\left(0\right)=0$, and $\dot{r}\left( 0 \right)=1$
(from Eq.~\ref{dotr}). Our goal is to search for solutions which
have two zeros of $r$ with $\dot{x}=0$ at both and $x$ evolving
monotonically between these turning points. For a given potential,
there is possibly one such solution, uniquely determined by the
initial position $x_{I}$. It is this quantity that we are ultimately
solving for.

We have implemented this search numerically, and have preliminary
results for the regime of large $\epsilon$ (order one)
\begin{Footnote} Further details of the numerics, including a
Mathematica notebook containing some sample calculations, can be
found at:  {\verb http://physics.ucsc.edu/~mjohnson/ .}
 \end{Footnote}. As
$\epsilon$ is decreased, the time scale between the zeros of $r$
increases, and since the field can only pause near the extrema of
the potential, we are forced to introduce an extreme fine tuning in
$x_{I}$ to induce this unnatural behavior. This is the fundamental
limitation on the numerics, but there is a regime where the fine
tuning in $x_{I}$ is reasonable (a few orders of magnitude less than
the machine precision). Shown in Fig.~\ref{xvss1} and \ref{rvss1} is
the evolution of $x(s)$ and $r(s)$ for $\epsilon=.6$ in the
potential Eq.~\ref{potential} with $b=.3$ and
$z=\{1,~.3,~.1,~.03,~.003\}$. These are the unique one-pass
solutions, and they satisfy all of the criteria outlined above. As
we take $z \rightarrow 0$, we find that the trajectories $x(s)$ and
$r(s)$ approach a constant profile (the blue curves). The reason for
this behavior can be seen in Fig.~\ref{x0vsz}, where we have plotted
$x_{I}$ as a function of $z$ for a few different values of
$\epsilon$. In each case, the initial position approaches a constant
as we take $z$ to zero, and since it is the initial position $x_{I}$
that uniquely determines a solution, we should expect to see the
trajectories $x(s)$ and $r(s)$ approach a constant profile.

\EPSFIGURE{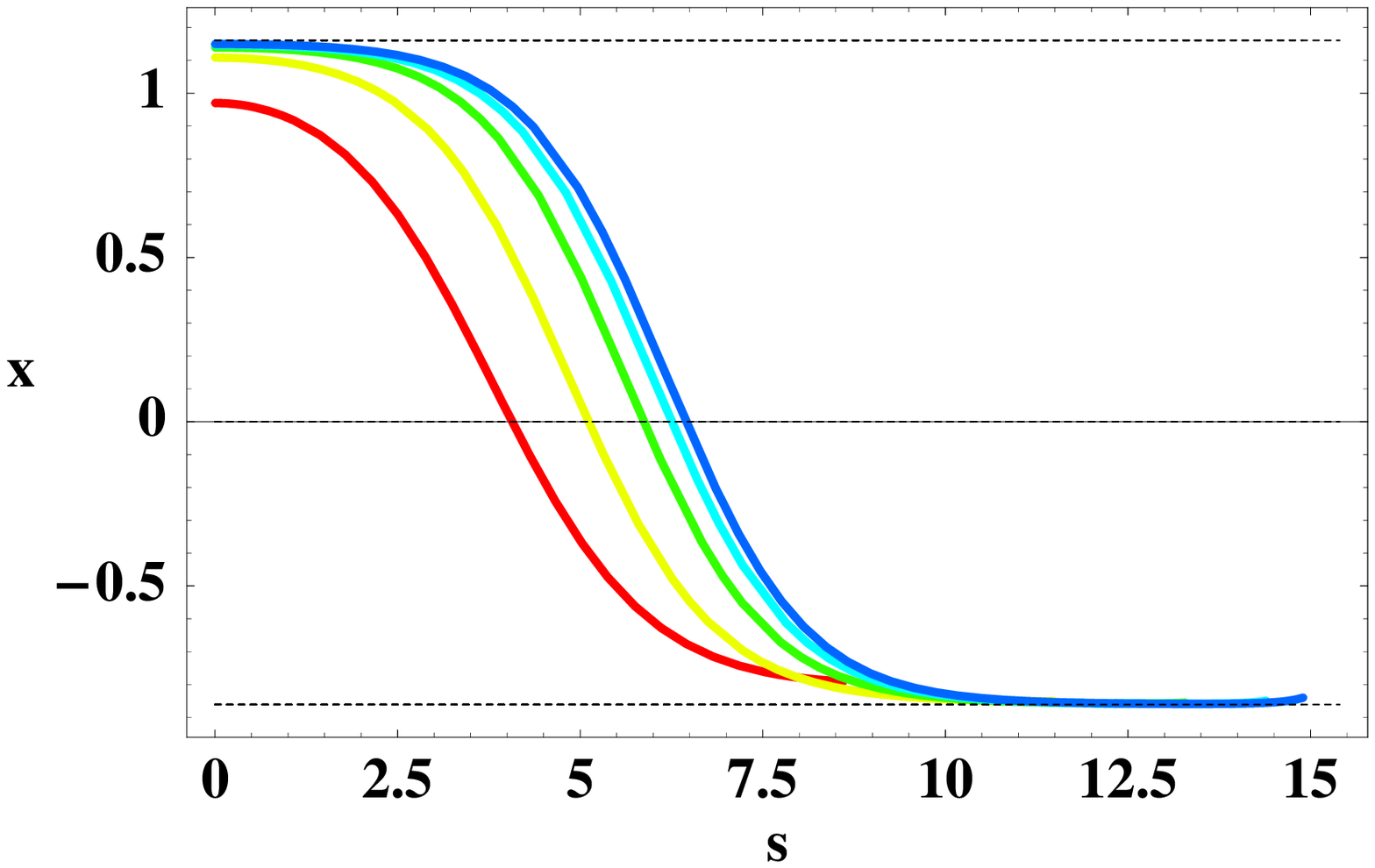,width=150mm}{The evolution of $x(s)$ in
the potential  Eq.~\ref{potential} with $b=.3$ for $z=\{1,~.3,~.1,~
.03,~.003\}$ and $\epsilon=.6$. The red curve is for $z=1$ and the
blue curve is for $z=.003$. Horizontal dotted lines represent
$x_{F}$ (bottom) and $x_{T}$ (top). \label{xvss1}}

\EPSFIGURE{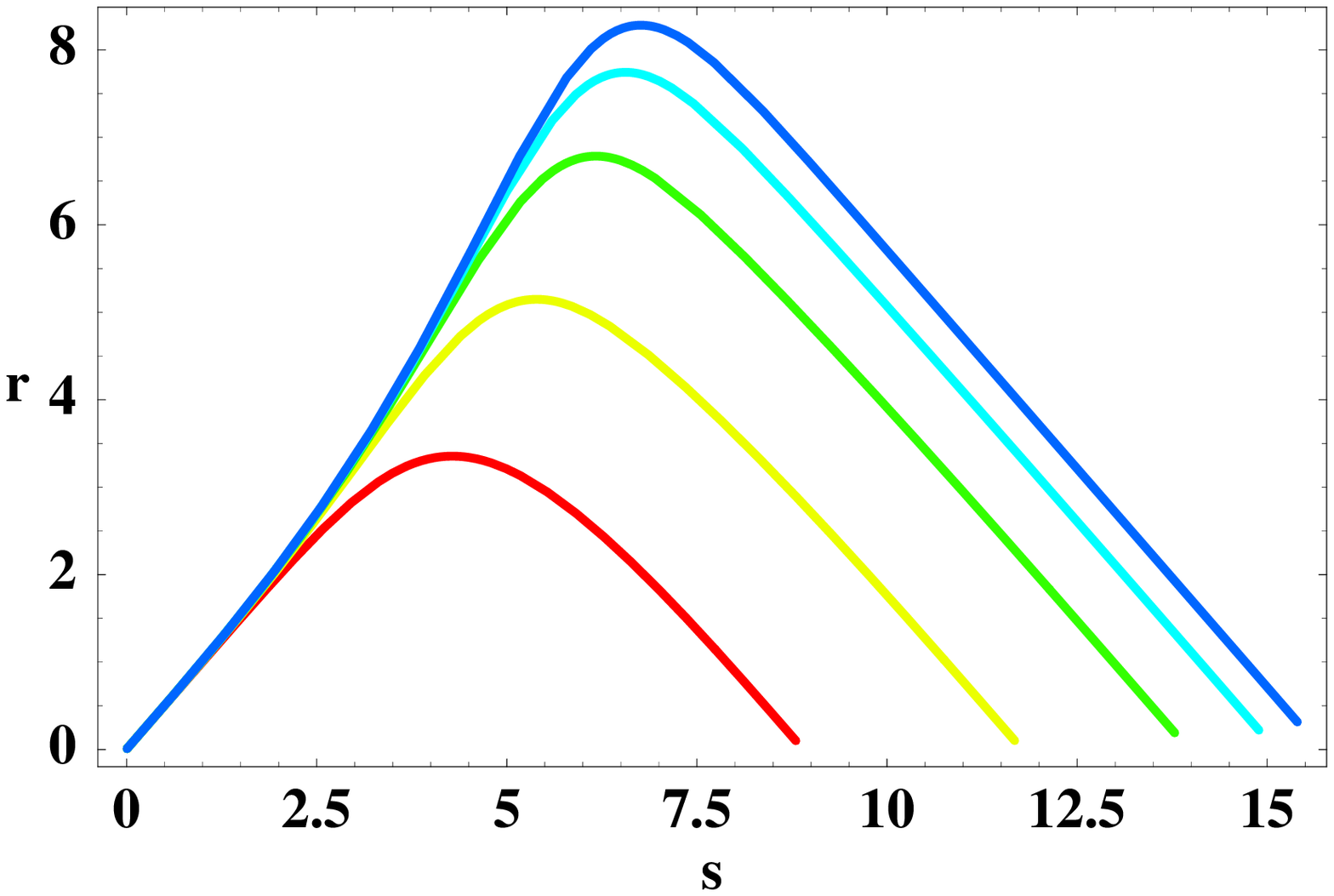,width=150mm}{The evolution of $r(s)$ in
the potential  Eq.~\ref{potential} with $b=.3$ for
$z=\{1,~.3,~.1,~.03,~.003\}$ and $\epsilon=.6$. The red curve is for
$z=1$ and the blue curve is for $z=.003$. \label{rvss1}}

Another observation we make is that as $\epsilon$ decreases, the
initial position (both for low- and high-z) approaches $x_{T}$ (the
horizontal dotted line in Fig.~\ref{x0vsz}). The time scale over
which $r$ evolves between its two zeros will increase as $\epsilon$
decreases (this can easily be seen in Eq.~\ref{dotr} and
\ref{ddotr}). We described in section~\ref{dstransitions} that $x$
can only loiter in the neighborhood of the positive  extrema of the
potential, and therefore as $\epsilon$ is decreased the final field
value will approach the position of the false vacuum minimum.
Simultaneously, to avoid under-shooting, we must adjust the initial
position closer to the position of the true vacuum minimum. As we
observed in section~\ref{dstransitions}, this corresponds to a
decreasing Gibbons-Hawking temperature in the false vacuum as
$\epsilon \rightarrow 0$.

\EPSFIGURE{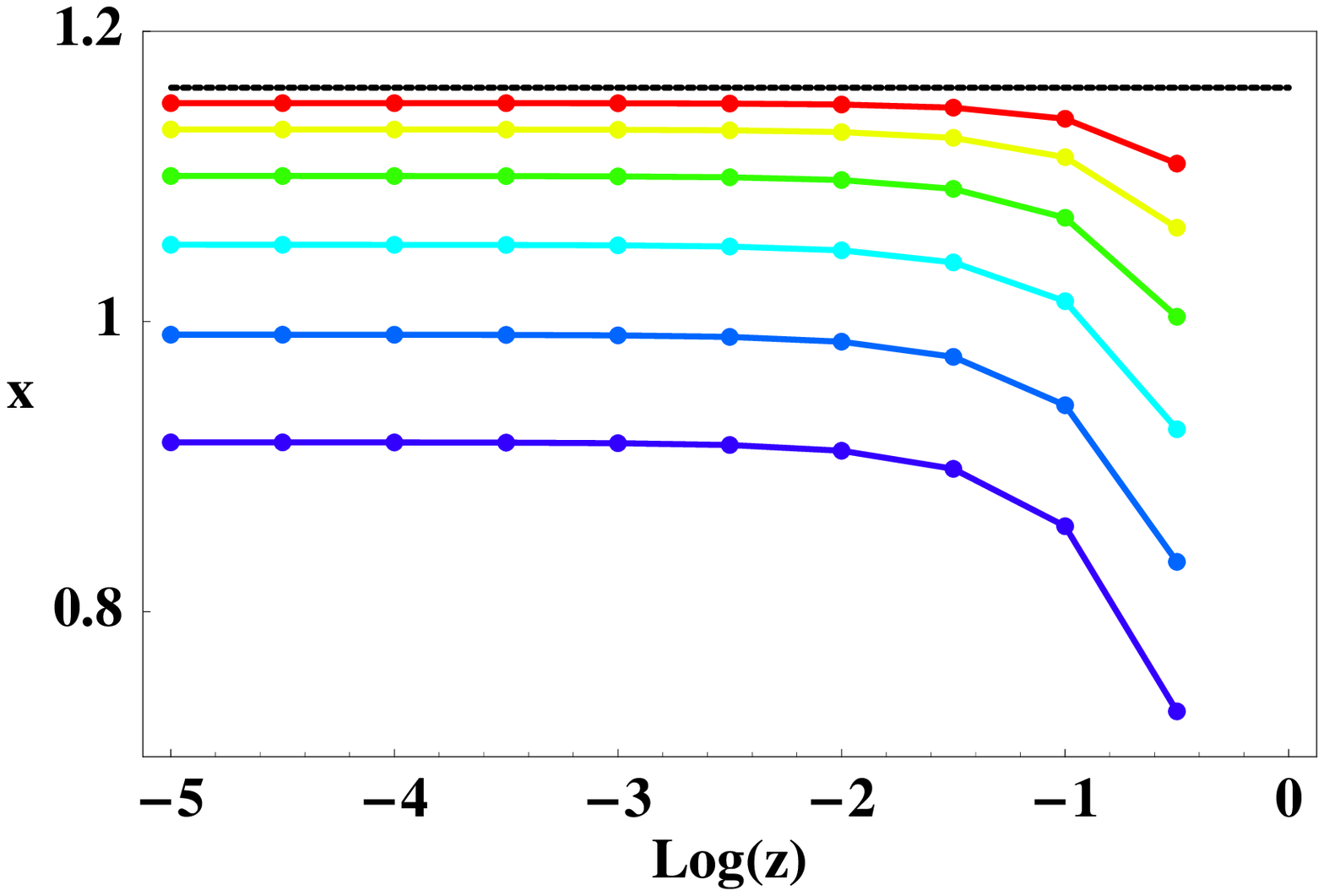,width=150mm}{The initial position of the
instanton solution, $x_{I}$,  vs $Log_{10} (z)$ for $.6 < \epsilon <
.85$ (red curve is for $\epsilon = .6$, violet curve is for
$\epsilon = .85$.) in the potential Eq.~\ref{potential} with $b=.3$.
\label{x0vsz}}

We claimed in section ~\ref{crunch} that the volume of the instanton
in the false vacuum phase will {\em not} increase indefinitely as
$V_{F} \rightarrow 0$, and therefore there is {\em no} regular
instanton which approaches the flat space solution in this limit.
The  result that $r(s)$ and $x(s)$ approach constant trajectories
confirms this conjecture, because the time the solution spends near
$x_{F}$ will become independent of $V_{F} $ after $z$ goes below
some critical value. Calculating the Euclidean action of these
solutions, given by Eq.~\ref{action}, we see in Fig.~\ref{Svsz} that
the instanton action asymptotes to a constant value. This shows that
the instanton action does not keep pace with the background
subtraction term, which is diverging as we take $V_{F} \rightarrow
0$. We therefore conclude that the probability to transition to a
big crunch spacetime, given by Eq.~\ref{Pft}, will not approach the
flat space answer, but instead will exponentially approach zero.

\EPSFIGURE{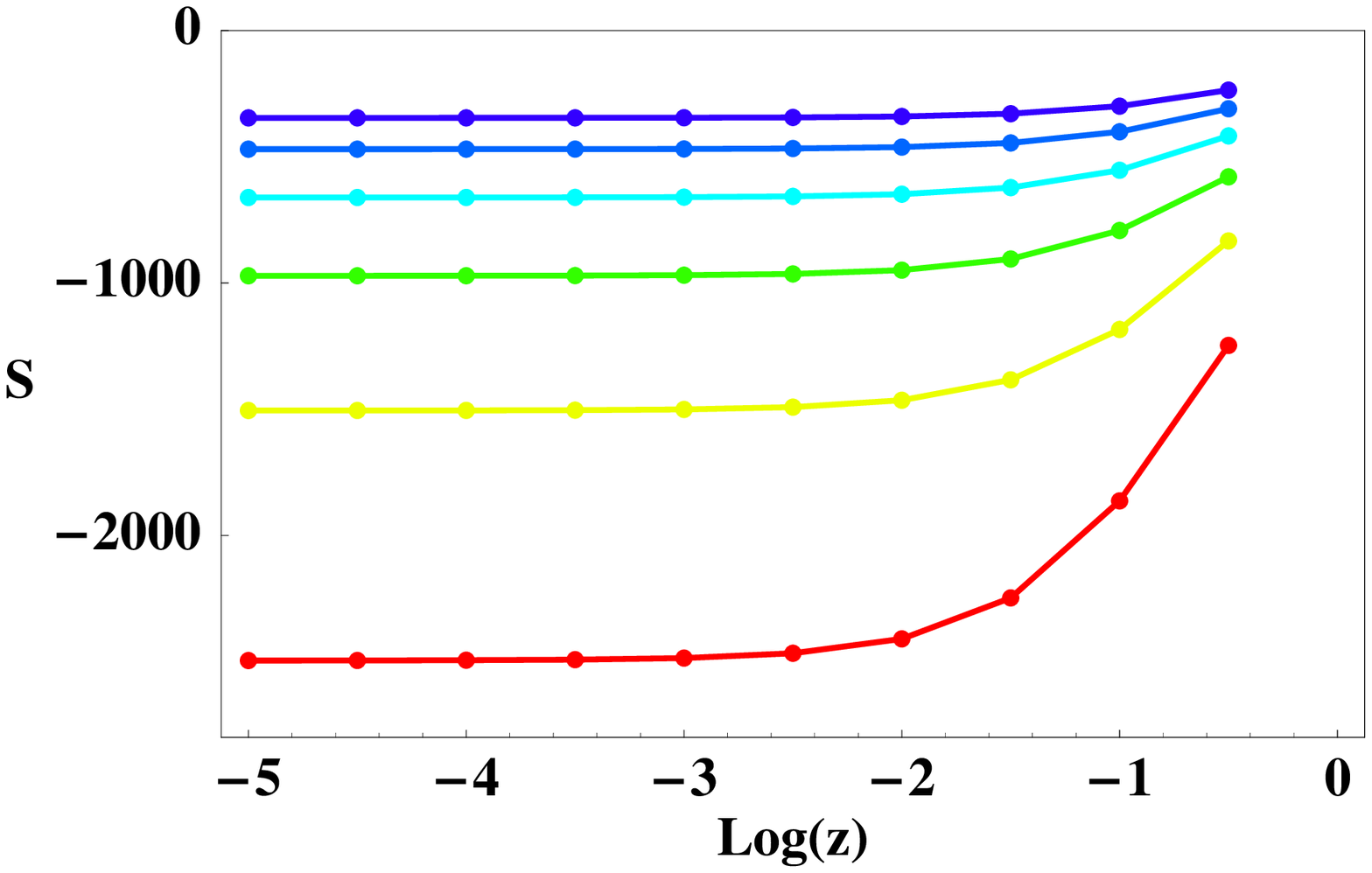,width=150mm}{The Instanton action (normalized to
$\left(\frac{M}{\mu}\right)^4$) vs $Log_{10} (z)$ for $.6 < \epsilon
< .85$ (red curve is for $\epsilon = .6$, violet curve is for
$\epsilon = .85$.) in the potential Eq.~\ref{potential} with $b=.3$.
\label{Svsz}}

We have explored other potentials, running over the range in $b$
shown in Fig.~\ref{potentials}, and found the same qualitative
behavior in all cases. We have also found the oscillating bounce
solutions \cite{heretics,hackworth} in each potential, and
explicitly verified that they have a higher total Euclidean action
(including the background subtraction) than the one-pass solutions
for all $z$. This numerical study will be extended to include
smaller values of $\epsilon$ in future work \cite{abj}, but we feel
confident that the preliminary results presented above are quite
generic.

\section{Tunneling to a state with vanishing vacuum energy}

The case of dS space tunneling into a space-time with vanishing c.c.
is a limit of the cases we have studied, and like all limits, must
be approached with care.  We will let $R$ stand for the radius, in
Planck units, of the dS space whose c.c. is being taken to zero. Let
us recall that in this case, as in that of a true vacuum with
negative c.c., it is inaccurate to state that we approach flat
space, even if the field asymptotes to a finite constant value at a
zero energy minimum of the potential. The local geometry indeed
approaches that of flat space-time, but the asymptotics are quite
different.  The space-time does have a null future conformal
boundary, where we can imagine organizing the states of the system
as ``multi-particle scattering states". However, it was argued in
\cite{landskept} that generic scattering data extrapolate back, not
to the smooth CDL space-time obtained from analytic continuation of
the instanton, but to a space-like singularity.    In asymptotically
flat space-time (which we assume to be exactly
supersymmetric\cite{tbfolly}) generic scattering data lead to at
worst localized singularities, perhaps hidden behind event horizons
(the Cosmic Censorship Hypothesis), which eventually give rise to
(or evolved from) other scattering data.

This is to be contrasted with the Lorentzian CDL instanton for $dS
\rightarrow dS$ transitions.   Here also the asymptotic geometry is
different from that of dS space, but in that case the global
geometry is irrelevant to any observation.  Causal diamonds remain
finite in area and the geometry inside any causal diamond approaches
that of the static patch of dS space at late times.  There are no
analogs of the catastrophic scattering data in these space-times
{\it if one makes the hypothesis that dS space has a finite number
of states.}   That is, in exact global dS space, using the formalism
of quantum field theory in curved space-time, we can construct
``scattering states" at past (future) null infinity which evolve to
future (past) Big Crunch (Bang) singularities.   According to the
conjecture of \cite{tbfolly}\cite{willy} these scattering data are
not supposed to be part of the quantum theory of gravity in dS
space, which instead one should imagine to be the quantization of
the phase space of gravitational solutions which are globally dS in
both the past and the future.   Similarly, in the CDL instanton for
dS transitions, we reject perturbations of the future asymptotic
data, which extrapolate back to Big Bang singularities. This allows
all excitations which can be seen inside a dS causal diamond, but
not black holes larger than the dS radius.  Thus, the structure of
CDL instantons for dS transitions fits well with the hypothesis that
dS space has a finite number of states.

Now imagine taking a limit in which one of the positive potential
energy densities is taken to zero.   Referring again to what happens
with the limit of pure dS space to flat space-time, we know that we
will have to throw away many of the quantum states of the theory in
order to have a sensible limit.  The entropy of the dS theory at
radius $R$ in Planck units is of order $R^2$, but the Poincare
invariant limit keeps only of order $e^{R^{3/2}}$ states. The
question is, {\it which states do we keep in a theory of CDL
tunneling from dS space to negatively curved FRW spaces with
vanishing c.c.?}

It seems to us that there are two reasonable hypotheses to make, in
answer to this question.  The first, which corresponds to what
passes for conventional wisdom about this problem, was made
in\cite{lenben} \cite{landskept}: admit the analog of all the
(future) scattering states we would keep in taking the limit of dS
space to Minkowski space.   The result, if it exists, is a cosmology
with an infinite number of states.  The initial states would all
correspond to a singular Big Bang, and most of them would transition
to the final FRW state without passing through an intermediate dS
phase.

One should also note the proposal of Susskind and
Freivogel\cite{lenben} in which one attempts to interpret the time
symmetric Lorentzian CDL instanton as the arena for a scattering
theory.  Given the observations of \cite{landskept} it seems to us
that this proposal has properties similar to the Big Bang
interpretation. It has the added problem of explaining how the Big
Crunch which follows from generic past asymptotic data evolves to
the Big Bang which preceded generic future asymptotic data.

We think there may be another interpretation of this space-time
which makes sense.   For very large but finite $R$ we can construct
scattering experiments with greater and greater precision in the
causal diamonds of the large radius bubble. Following the
prescription of \cite{tbds}, the theory contains an operator ,
$P_0$, whose low lying eigenvalues are approximately conserved. This
conservation law gets better and better as $R \rightarrow \infty$.
We can use it to restrict the possible black holes that are produced
in scattering experiments inside the causal diamond, to those whose
radius is small enough that they fit inside the CDL geometry without
causing a Crunch of the entire space-time. Since $P_0$ becomes a
conservation law as $R$ goes to infinity, and the horizon states
that give rise to thermal fluctuations decouple, we may expect the
S-matrix to be unitary in this subspace.   If this were the case, we
would have a consistent definition of scattering theory in the time
symmetric CDL space-time, involving only a finite number of in and
out states.  Despite the finiteness of the limiting system we could
hope that the scattering matrix was mathematically well defined.  In
effect, the huge set of localizable states in the large R
limit\footnote{Note that we are proposing to throw most of these
localizable states away in the CDL instanton, rather than keeping
them as we would in the flat space limit of dS space.}, would act as
classical measuring devices for the finite limiting system defining
the theory in the CDL instanton background. We will see below that,
from the point of view of the Landscape of String Theory, there are
advantages to a definition of the CDL instanton space-time, which
has only a finite number of states. Note however that in the stringy
context there is no way to consider the non-accelerating FRW
cosmologies, which arise from CDL instantons, as limits of
situations with positive c.c. The Dine-Seiberg regions of moduli
space, where these cosmologies find their home, are non-compact and
asymptotically supersymmetric. It would contradict the perturbative
assumptions on which the string landscape is based, to postulate a
positive c.c. in these asymptotic regions.

We want to emphasize that there is no positive evidence that the
prescription of keeping only a finite number of states is a
consistent one.   The problem is that effective field theory in a
CDL instanton background is not a self consistent approximation.
Innocuous looking perturbations of the future asymptotic data
evolved from situations in the past where effective field theory
breaks down.   We now have at least three proposals for the nature
of the theory of these systems: Big Bang, Freivogel-Susskind
scattering theory, and the finiteness hypothesis we have just put
forward, all of which might be wrong.

\section{Applications}
\subsection{Eternal inflation}

We have proposed an entirely new interpretation of the CDL tunneling
processes which are usually thought to lead to eternal inflation.
This interpretation is holographic and never discusses regions
outside the causal horizon of a given observer.   According to this
interpretation CDL tunneling only makes sense when the ``false
vacuum" has non-negative (perhaps only positive) cosmological
constant\footnote{The point is that AdS minima, with the usual
boundary conditions on fluctuations, do not allow instanton
solutions. In some cases \cite{hh} it has been shown that the CDL
instanton corresponds to a boundary field theory with a Hamiltonian
which is not bounded from below.  It may be possible to stablize
these theories, but it is unlikely that the stable ground state has
an interpretation in terms of low curvature bulk space-time.}. In
the case where both minima have positive c.c., the physics is
interpreted in the finite dimensional Hilbert space of the dS space
with smaller c.c. The instanton splits space-time up into two
causally disconnected regions, one of which relaxes back to the
true, and one to the false vacuum.   The false vacuum is interpreted
as a low entropy meta-stable state of the high entropy true vacuum
system. The word vacuum is a bit of a misnomer because both states
are relatively high entropy, low temperature, thermal systems.   The
instanton describes tunneling back and forth between these states,
and the rates are related by the principle of detailed balance. The
latter fact is a piece of semi-classical evidence that the current
interpretation of instanton physics is the correct one. This
interpretation implies that for the case of two positive
cosmological constants, there is no sense in which the ``false
vacuum dominates the global physics of the universe". That
conventional claim is based on attributing independent degrees of
freedom to every horizon volume of dS space, into the infinite
future in global time.   That hypothesis is rejected by the
holographic interpretation, and, we claim, at least weakly
disfavored by the form of the semi-classical tunneling amplitudes.
We do not see how to interpret the detailed balance result in terms
of the conventional picture.

The case where the ``true vacuum" has negative c.c., much larger in
absolute value than the positive c.c. of the ``false vacuum", was
studied in this paper.   In this case it {\it is } true that the
false vacuum dominates the physics of the system, but the details
are very different than the conventional interpretation would lead
us to expect.   The situation is again one of a high entropy system
(the false vacuum) making rare transitions into a low entropy
meta-stable state.   However, the false vacuum dominates by virtue
of its large entropy, rather than its large expansion rate, so the
effect is most marked for very large dS radius, rather than small.
This introduces a puzzling discontinuity with the calculation for
decay of flat space into the same negative c.c. crunch.   We
explained that this was due to the very different nature of the
asymptotically flat and dS ``vacuum states".   The dS vacuum is a
highly degenerate mixed state of microscopic degrees of freedom
which are mostly beyond the ken of local observers.   The flat limit
is taken by discarding most of these states and retaining only a set
of states localized in a single horizon volume, which become stable
in the $R \rightarrow\infty$ limit.   They are graded by a new
conserved generator, the Poincare Hamiltonian, whose commutator with
the static dS Hamiltonian vanishes in this localized subspace in the
limit.   The flat space vacuum is the non-degenerate zero-eigenstate
of the Poincare Hamiltonian.   Its decay rate is unaffected by the
entropy considerations that make the decay of the dS vacuum
improbable, and it is this discontinuity, which is reflected in the
CDL calculations.

We should note that if the hypothesis of Cosmological SUSY breaking
is correct, then the discontinuity disappears.   This hypothesis
states that the limit of dS space as the cosmological constant goes
to zero is exactly supersymmetric and R symmetric.   There is no CDL
instanton for the decay of such a supersymmetric state, and
supersymmetry leads to a positive energy theorem\cite{wsy}, which
implies that the vacuum is stable.

Finally, we recall that there is an entirely different mechanism
which is usually covered by the rubric of eternal inflation.  This
is the {\it self-reproducing inflationary universe} of Linde\cite{l}
(see also earlier work of\cite{starobinsky}). Our considerations
have nothing explicit to say about these models. However, the
self-reproducing universe shares with the instanton based ideas of
eternal inflation, a mechanism which depends on the existence of
independent degrees of freedom in each horizon volume of dS space at
infinite future time.   Our reinterpretation of the CDL instanton
transitions denies the existence of this infinite set of degrees of
freedom.  Thus, it is philosophically (though not yet mathematically
or physically) contradicting the basis of self-reproduction as well.

\subsection{A toy landscape}

There are two extant proposals for incorporating the observational
fact of an accelerating universe into ``string theory".   The first
is the string theory landscape\cite{land} and the second the theory
of stable dS space\cite{tbds}.   Neither proposal is a fully
developed mathematical theory.

We will first examine the implications of our interpretation of CDL
instantons, for the landscape, under the (false) assumption that
there are no Dine-Seiberg regions with potential falling to zero at
infinite places in moduli space.  This is then an imaginary
landscape in which every minimum has either positive or negative
energy density.    For such a landscape, our interpretation of
tunneling phenomena leads to the conclusion that the ``stable ground
state" is the minimum with maximal area causal diamond in the
Lorentzian section of a CDL instanton, which has the field in the
basin of attraction of that minimum.  For dS minima, this maximal
area is the dS horizon area, while for negative c.c. Big Crunches
the area depends on more details of the potential.

The maximal area principle implicitly assumes that the landscape is
connected.   That is, for each minimum of the potential, we can make
a list of all other minima that can communicate with it through a
sequence of tunneling events.   For a general potential, this might
break the space of minima up into disconnected classes, and the
maximal area principle would refer to each class independently.

A slightly peculiar feature of this proposal would arise if it
turned out that the maximal area causal diamond occurred for
negative c.c.    In that case the life-time of observers (see the
footnoted definition above) in what we call the ground state would
be power law in the area.   By contrast, if the ground state has
positive c.c. an observer's life-time is an exponential of a power
of the area of the maximal causal diamond.   Thus, in the case of a
negative c.c. maximal diamond we would probably want to invent an
{\it observerphilic} principle\footnote{Unlike the anthropic
principle, this principle makes no reference to biology or life.
Rather it is a statement about the maximum lifetime of large quantum
systems that are well described by quantum field theory, in a system
where a typical localized state of high enough energy, is a black
hole. Crude estimates of life-times can be made using well
understood physics.}. The quantum system representing this
landscape, might spend most of its time in a Big Crunch ground
state, but the most likely situation to be observed (by an
exponentially large factor) would be the minimum with smallest
positive cosmological constant. Thus, the combination of the maximal
area and observerphilic principles predicts that we should expect to
see a positive cosmological constant with the smallest value allowed
by the landscape.   If these principles were applicable to the
string landscape, we would have to hope that the current estimates
of the number of meta-stable dS ground states is wrong and that the
real number is of order $10^{120}$. Furthermore, these principles
pick out a unique ground state, so all other features of the low
energy physics would be specified, with no recourse to the anthropic
principle. Weinberg's\cite{weinberg} argument would have to be
relegated to the category of interesting coincidences.    This sort
of vacuum selection principle was anticipated in \cite{tbsb}.

We should point out that the conclusion that an eternally inflating
system could choose the state of lowest positive c.c. has been
derived by Linde on a rather different basis.  In the stochastic
approach to inflation invented by Starobinsky\cite{starobinsky}, if
one looks for stationary solutions of the equations\cite{lcc} one
concludes that in a given horizon volume, the most probable value of
the c.c. is the smallest possible positive one.  However, within the
context of the global view of eternal inflation, Linde argues that
it is by no means clear that one should accept the need for
stationary solutions.   Followers of the holographic point of view,
who reject the existence of degrees of freedom extraneous to the
largest causal diamond of any observer in the system, find the
argument about stationary distributions more convincing.

\subsection{The real string landscape}

Of course, the real string landscape is not of the type described
above.   It has Dine-Seiberg regions in which the potential goes to
zero.   The CDL instanton geometry in the basin of attraction of
those regions describes a space-time which is locally more and more
like 10 or 11 dimensional flat space.  Supersymmetry is restored
asymptotically from the point of view of local physics, and weakly
coupled string theory, or low energy 11D SUGRA appears to describe
the local low energy physics. Causal diamonds of infinitely large
area exist in this space-time.

The conventional wisdom is that there is only one string landscape,
all of whose dS minima can tunnel (possibly via multiple steps) into
each other and into any of the three {\it asymptotically maximally
supersymmetric} (AMS) regions of moduli space.   We should emphasize
that while we have only weak evidence that any stringy landscape of
dS minima exists we have exactly zero evidence that there is only
one connected landscape.   We think that anyone who has thought
about this at all finds the prospect of three different kinds (IIA,
IIB, 11D) of asymptotic regions in the same theory confusing.   Is
the full Hilbert space a direct sum of Fock spaces of outgoing
particle states in these different space-times?   Or are they
``dual" to each other, whatever that means?   In order to avoid
having to think about these skull-bursting questions, one might hope
that the landscape actually consists of disconnected families of dS
minima, with each family tunneling into only one kind of AMS
universe.

In fact, if one accepts the conventional wisdom, there is a similar,
if slightly more subtle problem for any system where two or more dS
minima can tunnel into the {\it same} region in field space.  The
point is that there are then multiple instanton solutions.   The
Lorentzian continuation of each instanton contains a region of
negatively curved FRW universe, which asymptotes to a matter or
kinetic energy dominated cosmology.  Each FRW universe is slowly
varying in the future, so we can introduce an {\it adiabatic Fock
space}\cite{bd} for the quantization of field fluctuations around
the classical background.    However, because the backgrounds are
globally different on the homogeneous slices of constant negative
curvature, the Bogoliubov transformation between the two quantum
operator algebras is not implemented by a unitary transformation on
Fock space\footnote{Any two Hilbert spaces of the same (finite or
countable) dimension are related by (many) unitary transformations.
However, these transformations do not usually map simple observables
to each other.   The Bogoliubov transformation provides a simple map
of observables, but not a unitary map of Hilbert spaces.}.   So even
in the simplest model of a landscape we have to deal with the issue
of how to include globally different asymptotic space-times in the
same quantum theory.

This problem would be avoided if we could make sense of the
suggestion that only a finite number of scattering states in each
asymptotic region are actually allowed in the context of a
landscape.   The lack of unitarity of the Bogoliubov transformation
is related to the infinite number of degrees of freedom in quantum
field theory, and would certainly disappear if we could restrict
attention to a finite number of states.

While it is not clear that this is possible, the attraction of the
proposal is evident.   In particular, if this proposal makes sense,
the Stringy Landscape would be similar to the toy landscape of the
previous subsection and we could borrow the dynamical selection
principle from that discussion.   {\it The String Landscape would be
a system with a finite number of states\footnote{This assumes that
there are a finite number of meta-stable dS vacua, an assertion
which is now open to doubt.}. It could best be viewed from the point
of view of the meta-stable minimum with smallest positive
cosmological constant. Tunneling transitions would be interpreted in
terms of the quantum mechanics of the static Hamiltonian of this dS
space. Minima with larger absolute value of the cosmological
constant\footnote{At this point the reader should refer back to the
discussion of the observer-philic principle, to recall the extra
complication that ensues if the smallest absolute value actually
corresponds to a negative c.c.} are low entropy subsystems, which
the system visits rarely, and for periods brief compared to its
sojourns in the ``ground state".} This would make the string
landscape into a predictive model of low energy physics. From many
points of view, this conjecture puts the string landscape on the
same footing as the equally hypothetical theory of stable dS space.
The main difference is the attitude towards SUSY.   Part of the
hypothesis of \cite{tbfolly} is that the scale of SUSY breaking is
directly connected with the cosmological constant, through the
formula $m_{3/2} = c \Lambda^{1/4}$.  In the Landscape there does
not appear to be a similar connection between the value of the c.c.
and SUSY breaking.

\subsection{The theory of stable dS space}

The current paper actually arose out of the one of the authors'
(T.B.) consideration of low energy effective theories which could be
compatible with the hypothesis of Cosmological SUSY Breaking.   It
has proven extraordinarily difficult to find any low energy
implementation of CSB, which was also compatible with a reasonable
phenomenology. The only extant model looks peculiar from the
conventional standpoint of low energy SUSY theorists.   Once the
parameters have been set to accord with the ideas of CSB, the model
looks more or less like a conventional model of low energy dynamical
SUSY breaking. In particular, much of the dynamics of SUSY breaking
is determined by a non-gravitational model of gauge and chiral
superfields.   This model has, in addition to its SUSY violating
vacuum state, a moduli space of SUSic vacua.

In \cite{susycosmophenoIV} one of the authors (T.B.) made the
general observation that non-gravitational theories with SUSY
violating meta-stable minima, as well as SUSY vacuum states may be
useful for phenomenology. The point is that if we decide to fine
tune the c.c. so that it is small and positive at the SUSY violating
minimum, then the SUSic vacua have negative c.c. of microphysical
scale.   This means that they have NOTHING to do with the theory in
which the SUSY violating dS space lives.   The ``meta-stable" SUSY
violating vacuum decays into a SUSY violating Big Crunch cosmology,
not one of the SUSic vacuum states.   The question then becomes one
of the lifetime of the ``meta-stable" state.  This paper gives a
surprising answer to that question.  It is easy to construct low
energy models in which the instanton for decay of a flat, SUSY
violating meta-stable minimum, has an action so small that the decay
would occur on a time scale shorter than the current age of the
universe.   This paper shows that the flat space estimate is
irrelevant to the calculation of the true vacuum decay lifetime for
dS space, which will be superexponentially longer than the age of
the universe if the c.c. is tuned to be exponentially small.

Apart from the numerical/phenomenological significance of this
result, these considerations resolve an issue of principle for the
hypothesis of CSB.  Under that hypothesis, dS space is supposed to
be stable.   We argued that when the negative c.c. has microphysical
scale, the CDL process did not represent an instability.  Rather it
represents a low probability transition to a low entropy meta-stable
state, analogous to all the air in a large room collecting in a
small box with a pinhole in it, situated somewhere in the room.

\section{Conclusions}

The form of the Coleman-DeLuccia probabilities for tunneling to and
from a large radius dS space supports a holographic reinterpretation
of eternal inflation.   When the potential landscape is such that
the smallest absolute value of the c.c. occurs for positive c.c. one
considers the most stable state of the system to be the dS space
with that value of the c.c.   Other minima are considered as low
entropy meta-stable states of the system, which (for entropic
reasons) are visited infrequently, and out of which one tunnels with
relative rapidity.  For pairs of minima with positive c.c., the
success of the detailed balance prediction for the relative rates of
transition gives, in our opinion, strong support for this
interpretation.   The fact that the largest causal diamond in a
negative c.c. Big Crunch\footnote{T.B. cannot resist emphasizing for
the Nth time that the AdS solution of the field equations at the
negative c.c. minimum has nothing to do with any of the physics
discussed in this paper.   It is an isolated theory of quantum
gravity with AdS boundary conditions, completely described by a
conformal field theory.} has area of microphysical size helps to
explain the smallness of the rate for the ``decay" of the large
radius dS space into this state. It also motivates an estimate of
the (much larger) inverse transition rate from the Big Crunch to dS
space. These ideas also explain the puzzling discontinuity in the
CDL transition rate to a negative c.c. crunch, when the positive
c.c. is lowered to zero.

If the lowest absolute value of the c.c. occurs for negative c.c.,
the fundamental description of the physics is more obscure. Assuming
that it can be defined, our generalization of the CDL amplitudes
again suggest a finite entropy system, which spends most of its time
in the Big Crunch state.   However, in this case, and assuming that
there is a positive c.c. minimum of very small size, the question of
the lifetime of observers comes into play.   For our purposes, an
observer is a large, localized system, approximately described by
cutoff quantum field theory.  Such systems can survive in the dS
state, for times which are exponential in an inverse power of the
positive c.c.   In the Big Crunch state they survive a time power
law in the (absolute value of) the negative c.c.  The Big Crunch
state is, for most of its history, a state where all degrees of
freedom are maximally excited and interacting.   The period when
observers can exist is short. Thus, even if the landscape has its
smallest c.c. negative, one predicts that a typical observer will be
found in the region with smallest positive c.c.   We emphasize again
that real theories of quantum gravity may only realize some (or
none) of the toy landscapes we are discussing here.

The most well motivated landscape scenario is based on string
theory.  Here, the potential has asymptotic Dine-Seiberg regions
where flat supersymmetric 10 or 11 dimensional space-time is locally
restored.   The possibility of regions with vanishing potential is
from the point of view presented here, a singular limit in which the
number of states of the system might become infinite.   The analysis
of stable dS space in \cite{tbds} indicates that such limits have to
be taken with great care.   Most of the states that exist for finite
dS radius, decouple and must be thrown away when considering the
Poincare invariant limiting theory.   A similar question arises for
landscapes with points in field space that have vanishing c.c.   We
presented several conjectural descriptions for such systems, none of
which is without problems.  We think that our mathematical results
on instantons are correct, and indicate the need for a
reinterpretation of eternal inflation.   Our suggestion that the
string landscape might refer to a quantum system with a finite
number of states is much more conjectural.

Our proposed reinterpretation of eternal inflation replaces an
infinite and incomprehensible\footnote{More precisely, ``as yet
uncomprehended"} fractal with a finite system, and suggests
different rules for deciding the probability of various observations
in an eternally inflating universe.   It may be, as advertised in
the title of this paper, that these rules can be viewed as a
regulated and invariant version of the ``volume weighted" counting
of probabilities, which has been suggested by the inventors of
eternal inflation.  In \cite{nightmare} the authors suggested that
the global gauge picture of de Sitter space might be approximately
justified in the limit of large dS space by the following stratagem:
In each horizon volume consider only those localizable states which
one would keep in the Poincare invariant limit, and treat them
(approximately) by the usual rules of particle physics.  The entropy
of such states scales as $(R M_P)^{3/2}$. The total dS entropy
allows us to have $(RM_P)^{1/2}$ commuting copies of these degrees
of freedom, which we can try to associate with disjoint horizon
volumes.   This would be compatible with the global gauge
description of dS space if we enforced an IR cutoff on the global
time.  Given a form for the potential, we can estimate the
probability for a given observer's Poincare ground state to nucleate
a bubble of Big Crunch, by the usual non-gravitational rules.  The
probability of finding dS thermal fluctuations of this state is very
much smaller than the local bubble nucleation probability, when the
dS radius is taken large with microphysical scales
fixed\footnote{Under the hypothesis of Cosmological SUSY Breaking,
some microphysical scales are actually tied to the c.c..} .

But what do we mean by a given observer in the above paragraph? The
arrangement of the states of dS space into states of a finite
collection of disjoint observers is a gauge choice with no direct
physical meaning.   All one observer can ever see are its own
localizable observables and a collection of degenerate degrees of
freedom which form a thermal bath for localizable observations.
There are not enough localizable degrees of freedom to measure the
precise state of the heat bath.    The only predictions of the
theory which should be taken seriously are those which refer to
averages over the different possible choices of which degrees of
freedom we choose to represent the ones that we actually experience.
We claim that the typical observer, in this sense, does not
experience a vacuum decay with the probability determined by the
approximately Poincare invariant calculation in a single horizon
volume.   Ultimately, the argument for this is the large discrepancy
in the maximal entropy allowed by the covariant entropy bound in the
two bubbles of the CDL instanton.   So the claim is that the
mathematically unambiguous CDL calculation of the tunneling
probability is taking into account the averaging over different
horizon volumes in global dS space, that is invoked in conventional
discussions of eternal inflation.   Those considerations are
modified by taking into account the covariant entropy bound.   The
global description is cut off in such a way that the entropy
attributed to a given dS space never exceeds one quarter of the area
of its horizon.

 Apart from the
radical reinterpretation of some kinds of eternal inflation models,
the most significant impact of our results is on dynamical SUSY
breaking models. Traditionally, one studies a non-gravitational
model and insists that it has no SUSY ground states.   Our results
suggest that this is much too restrictive.  It is sufficient to find
a quantum field theory with a metastable SUSY violating ground
state. If we then fine tune the c.c. to be very small and positive
at the SUSY violating point in field space, then the SUSic ground
states are irrelevant.  The meta-stable SUSY violating state tunnels
to a Big Crunch, but this is a very improbable event, in a system
with a finite number of states.   The system spends most of its time
in the meta-stable SUSY violating state.

\section{Acknowledgments}

One of the authors (T.B.) acknowledges a conversation with two
gentlemen (Luigi and Salvatore) from somewhere south of Verona, who
convinced him to include certain of the references in \cite{FSB} and
\cite{rb} in the current paper. Helpful comments from,
N.~Arkani-Hamed, and R. Bousso are also acknowledged.  A.Linde
helped us to understand some of his earlier work on this subject,
and gave generously of his time and numerical expertise. We
particularly want to thank Anthony Aguirre, who carefully read and
debugged the manuscript and gave us lots of good advice. This
research was supported in part by DOE grant number
DE-FG03-92ER40689.




  %




\end{document}